\def\BradExtended{}
\documentclass[sigconf, nonacm, pdfa]{acmart}

\newcommand\vldbdoi{XX.XX/XXX.XX}
\newcommand\vldbpages{XXX-XXX}
\newcommand\vldbvolume{17}
\newcommand\vldbissue{11}
\newcommand\vldbyear{2024}
\newcommand\vldbauthors{\authors}
\newcommand\vldbtitle{\shorttitle}
\newcommand\vldbavailabilityurl{\url{https://github.com/mitdbg/brad}}
\ifdefined\BradVLDB
  \newcommand\vldbpagestyle{empty}
\else
  \newcommand\vldbpagestyle{plain} 
\fi
\usepackage[inline]{enumitem}
\usepackage{xcolor}
\newcommand{\thesystem}{BRAD}
\newcommand{\htap}{System H}
\newcommand{\company}{QuickFlix}

\newcommand{\ScaleDownBradSelfSavings}{6.0$\times$}
\newcommand{\ScaleDownBradHtapSavings}{4.6$\times$}
\newcommand{\ScaleDownBradARSavings}{13$\times$}

\newcommand{\ScaleUpTxnBradARSavings}{2.6$\times$}
\newcommand{\ScaleUpTxnBradHtapDiff}{1.1$\times$}

\newcommand{\SpecializedBradARSavings}{2.4$\times$}

\newcommand{\BRADDayLongCumulativeCostImprovement}{1.7$\times$}
\newcommand{\BRADDayLongInitialCostImprovement}{2.5$\times$}
\newcommand{\BRADDayLongPeakCostImprovement}{2.1$\times$}

\newcommand{\SLOBradSavingsOverAR}{4.2$\times$}
\newcommand{\SLOHtapSavingsOverBrad}{4.4$\times$}

\newcommand{\BRADRouting}{1.31$\times$}
\newcommand{\RedshiftOnlyRouting}{1.85$\times$}
\newcommand{\RTOnlyRouting}{1.34$\times$}
\newcommand{\RTWithOverheadRouting}{1.54$\times$}
\newcommand{\RandomRouting}{3.78$\times$}

\newcommand{\BRADTxnOverhead}{2.46~ms}
\newcommand{\BRADAnaOverhead}{10~ms}

\newcommand{\BradBestCostSavings}{\ScaleDownBradARSavings}
\newcommand{\BradBestCostSavingsHtap}{\ScaleDownBradHtapSavings}
\newcommand{\BradBestCostSavingsAR}{\ScaleDownBradARSavings}
\newcommand{\BRADCostSavingsRange}{1.6--\BradBestCostSavings{}}

\newcommand{\sparagraph}[1]{\vspace{1mm}\noindent {\bf #1}}

\usepackage{overpic}
\usepackage{tikz}
\newcommand*\captionlabel[1]{%
  \tikz[baseline=(char.base)]{%
    \node[shape=circle,fill=white,draw=black,inner sep=1.5pt, outer sep=5pt] (char) {%
      \textcolor{black}{\scriptsize \textsf{#1}}};
}}
\newcommand\encircle[1]{%
  \tikz[baseline=(X.base)]
    \node (X) [draw, shape=circle, inner sep=-1pt] {\strut\resizebox{.75\width}{!}{#1}};}

\usepackage{tabularx}
\usepackage{algorithm}
\usepackage{algpseudocodex}

\usepackage{pbalance}

\begin{document}

\author{Geoffrey X. Yu}
\affiliation{%
  \institution{MIT CSAIL}
  \city{Cambridge}
  \state{MA}
}
\email{geoffxy@mit.edu}

\author{Ziniu Wu}
\affiliation{%
  \institution{MIT CSAIL}
  \city{Cambridge}
  \state{MA}
}
\email{ziniuw@mit.edu}

\author{Ferdi Kossmann}
\affiliation{%
  \institution{MIT CSAIL}
  \city{Cambridge}
  \state{MA}
}
\email{kossmann@mit.edu}

\author{Tianyu Li}
\affiliation{%
  \institution{MIT CSAIL}
  \city{Cambridge}
  \state{MA}
}
\email{litianyu@mit.edu}

\author{Markos Markakis}
\affiliation{%
  \institution{MIT CSAIL}
  \city{Cambridge}
  \state{MA}
}
\email{markakis@mit.edu}

\author{Amadou Ngom}
\affiliation{%
  \institution{MIT CSAIL}
  \city{Cambridge}
  \state{MA}
}
\email{ngom@mit.edu}

\author{Samuel Madden}
\affiliation{%
  \institution{MIT CSAIL}
  \city{Cambridge}
  \state{MA}
}
\email{madden@csail.mit.edu}

\author{Tim Kraska}
\affiliation{%
  \institution{MIT CSAIL, AWS}
  \city{Cambridge}
  \state{MA}
}
\email{kraska@mit.edu}

\newcommand{\PaperTitle}{Blueprinting the Cloud: Unifying and Automatically Optimizing Cloud Data
  Infrastructures with \thesystem{}}

\ifdefined\BradExtended
\title{\PaperTitle{} -- Extended Version}
\else
\title{\PaperTitle}
\fi

\begin{abstract}
  Modern organizations manage their data with a wide variety of specialized
  cloud database engines (e.g., Aurora, BigQuery, etc.).
  However, designing and managing such infrastructures is hard.
  Developers must consider many possible designs with non-obvious performance
  consequences; moreover, current software abstractions tightly couple
  applications to specific systems (e.g., with engine-specific clients), making
  it difficult to change after initial deployment.
  A better solution would \emph{virtualize} cloud data management, allowing
  developers to declaratively specify their workload requirements and rely on
  automated solutions to design and manage the physical realization.
  In this paper, we present a technique called \emph{blueprint planning} that
  achieves this vision.
  The key idea is to project data infrastructure design decisions into a unified
  design space (blueprints).
  We then systematically search over candidate blueprints using cost-based
  optimization, leveraging learned models to predict the utility of a blueprint
  on the workload.
  We use this technique to build \thesystem{}, the first cloud data
  virtualization system.
  \thesystem{} users issue queries to a single SQL interface that can be backed
  by multiple cloud database services.
  \thesystem{} automatically selects the most suitable engine for each query,
  provisions and manages resources to minimize costs, and evolves the
  infrastructure to adapt to workload shifts.
  Our evaluation shows that \thesystem{} meet user-defined performance targets
  and improve cost-savings by \BRADCostSavingsRange{} compared to serverless
  auto-scaling or HTAP systems.
\end{abstract}

\maketitle

\pagestyle{\vldbpagestyle}
\ifdefined\BradVLDB
  \begingroup\small\noindent\raggedright\textbf{PVLDB Reference Format:}\\
  \vldbauthors. \vldbtitle. PVLDB, \vldbvolume(\vldbissue): \vldbpages, \vldbyear.\\
  \href{https://doi.org/\vldbdoi}{doi:\vldbdoi}
  \endgroup
  \begingroup
  \renewcommand\thefootnote{}\footnote{\noindent
  This work is licensed under the Creative Commons BY-NC-ND 4.0 International License. Visit \url{https://creativecommons.org/licenses/by-nc-nd/4.0/} to view a copy of this license. For any use beyond those covered by this license, obtain permission by emailing \href{mailto:info@vldb.org}{info@vldb.org}. Copyright is held by the owner/author(s). Publication rights licensed to the VLDB Endowment. \\
  \raggedright Proceedings of the VLDB Endowment, Vol. \vldbvolume, No. \vldbissue\ %
  ISSN 2150-8097. \\
  \href{https://doi.org/\vldbdoi}{doi:\vldbdoi} \\
  }\addtocounter{footnote}{-1}\endgroup
\fi

\ifdefined\BradVLDB
  \ifdefempty{\vldbavailabilityurl}{}{
  \vspace{.3cm}
  \begingroup\small\noindent\raggedright\textbf{PVLDB Artifact Availability:}\\
  The source code, data, and/or other artifacts have been made available at \vldbavailabilityurl.
  \endgroup
  }
\fi

\section{Introduction}\label{sec:introduction}
Over the past decade, the cloud has transformed how organizations manage their
data through two key forces:
\begin{enumerate*}[label=(\roman*)]
  \item offering a plethora of specialized database engines optimized for diverse
  workloads~\cite{aws-redshift,aws-aurora,aws-athena}, and
  \item enabling ``one-click'' on-demand access to conceptually ``infinite''
  resources~\cite{ec2,s3,gcp,azure}.
\end{enumerate*}
To reap these benefits, cloud users must curate a collection of such
specialized database engines, each offering a competitive edge on different parts of
their workload.
For example, an organization might use Aurora~\cite{aws-aurora} to manage 
client accounts with transactions, Snowflake~\cite{snowflake} to analyze
historical sales data, and BigQuery Omni~\cite{bigqueryomni} for exploratory
analysis. 
Benefits aside, these multi-system infrastructures introduce
new management challenges.
Data engineers need to 
\begin{enumerate*}[label=(\roman*)]
  \item choose a suitable set of engines (out of
    dozens~\cite{aws-analytics,aws-db,gcp-db}) for their workload,
  \item partition and/or replicate their data across the engines,
  \item decide which engines to use for each aspect of their workload (i.e.,
    which queries go to each engine),
  \item provision the engines appropriately, and
  \item repeat these steps each time their workload or business needs change.
\end{enumerate*}
Navigating these decisions is hard; prior work showed that an optimal
infrastructure depends on many interconnected factors such as query selectivity,
service level objectives (SLOs), and dynamic load of the
system~\cite{brad-kraska23}.
Designs based on conventional wisdom can miss out on significant
performance and cost savings 
(Section~\ref{sec:motivational-examples}).
As a result, organizations struggle to design their infrastructure
while also keeping costs under control~\cite{gartner-cloudcost}.

To address this challenge, we recently presented our vision for
\thesystem{}~\cite{brad-kraska23}.
\thesystem{} is fundamentally a virtualization layer for cloud data
infrastructure.
\thesystem{} users do not specify the mapping of data to specific engines or
explicitly provision resources.
Instead, \thesystem{} uses a proxy-like indirection
layer~\cite{tigger-butrovich23, rds-proxy, pgbouncer} to abstract away multiple
database engines, appearing to end-users as a single SQL endpoint. 
Under the covers, \thesystem{} allocates data and operates the infrastructure
by picking the ``best'' set of engines for the workload, choosing the
appropriate data distribution and provisioning for each engine, and routing
queries optimally.
This is a fundamentally challenging because \thesystem{} must explore a huge
space of possible solutions, while meeting performance expectations.

We solve this problem using a novel technique we call \emph{blueprint planning},
which is a holistic \emph{cost-based optimization} over the infrastructure
design space.
Specifically, blueprints are system plans that define a \thesystem{} deployment.
They contain
\begin{enumerate*}[label=(\roman*)]
  \item the set of engines to include in the infrastructure,
  \item their provisioning configurations (e.g., instance type and number of
    nodes),
  \item the engine(s) on which each table in the dataset is placed, and
  \item a policy for routing queries to the engines.
\end{enumerate*}
Blueprints allow us to systematically and quantitatively consider all aspects of
the infrastructure design problem in a unified search space, analogous to
traditional query planning~\cite{selingeropt-selinger79}.
However, accurately assigning scores to blueprints is significantly harder than
query planning. First, the utility of a blueprint is not
captured by performance alone, as a good blueprint for a given workload
minimizes dollar-based operating costs under a latency-based performance
constraint (or vice-versa, depending on user-specified goals).
Second, accurately predicting a workload's performance (e.g., a query's run
time) on a blueprint is difficult due to (i) engines having opaque system
implementations and (ii) new constraints in our setting. 
Specifically, we must make these predictions when a physical query plan is
unavailable, preventing us from reusing existing learned models~\cite{SunL19,
MarcusP19, HilprechtB22, WuYYZHLLZZ22, marcus2019neo, marcus2022bao}.
For example, a candidate blueprint may add an engine into the infrastructure
that is not yet running (e.g., starting up a data warehouse) or replicate a
table onto a new engine to support a query.

In this paper, we show that these challenges are tractable.
In the cloud setting, infrastructure operators can collect performance data over
a wealth of workloads and deployments to build learned performance models.
Moreover, most query optimizers are deterministic. Thus, we can train a model to
predict a query's run time using just its logical properties (e.g., filter
selectivities, join templates) since the optimizer will pick similar query plans
with comparable run times for similar queries.
We leverage these observations to build a graph neural network with a novel
query featurization that relies only on such logical query features
(Section~\ref{sec:scoring-key-idea}).
Together with other analytical models, we use this model to predict the
performance and cost of candidate blueprints on a given workload.
We then use these predictions to drive a greedy beam-based search over the
blueprint search space to find an optimized infrastructure design.
We have implemented our blueprint planner in \thesystem{}, enabling it to
automatically design infrastructures consisting of three engines that cover a
large part of enterprise needs:
\begin{enumerate*}[label=(\roman*)]
  \item a transactional store (Aurora~\cite{aws-aurora}),
  \item a data warehouse (Redshift~\cite{aws-redshift}), and
  \item a data lake query engine (Athena~\cite{aws-athena}).
\end{enumerate*}

While there is a wealth of prior work in automatically optimizing single
systems~\cite{lim22, pavlo17, pavlo19, pavlo21, keebopaper, AutoWLM,
venkataraman2016ernest, ortiz2018slaorchestrator, ortiz2016perfenforce,
ortiz2019performance}, and in managing existing multi-engine
deployments~\cite{duggan15bigdawg, wang2017myria, alotaibi2019towards,
breitbart1992overview, breitbart1988update, hwang1994myriad,
pu1988superdatabases, sheth1990federated, georgakopoulos1991multidatabase,
josifovski2002garlic, bent2008dynamic, zhang2022skeena}, \thesystem{}
holistically automates and optimizes the design and operation of multi-engine
infrastructures.
Doing so involves reasoning about cost and performance across engines and
hypothetical deployments, which, to our knowledge, have not been studied.

We evaluate \thesystem{} by having it automatically optimize a data
infrastructure for cost under a performance constraint, 
We use a workload with both transactions and diverse analytics running on an
adapted version of the IMDB dataset~\cite{leis2015good}.
Overall, we show that \thesystem{} is able to react to changing workloads and
select designs that achieve performance targets in diverse deployment scenarios.
When compared to a baseline that na\"ively auto-scales transactional and
analytical systems, \thesystem{} achieves \BRADCostSavingsRange{} cost savings
due to its ability to route queries between engines and precisely scale to the
resource needs of a workload, instead of reacting passively to increased system
load.

\noindent
\ifdefined\BradExtended
\ \\[-0.7em]
\fi
\textbf{Contributions.} In summary, we make the following contributions:
\begin{itemize}[leftmargin=*]
    \item We introduce \emph{blueprint planning}: a new framework for
    virtualized, automated cloud data infrastructure design and management that
    applies cost-based optimization.

    \item We present a practical blueprint planning solution. We leverage a
    graph neural network with a novel logical query featurization that
    generalizes to common gradual workload changes.

    \item We present the design, implementation, and evaluation of \thesystem{}:
    a virtualized cloud database management system that uses blueprint
    planning to automate infrastructure design.
\end{itemize}

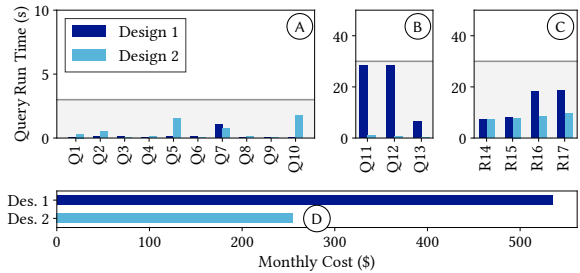
\begin{figure}[t]
    \vspace{-0.2em}
    \begin{overpic}[width=0.95\columnwidth]{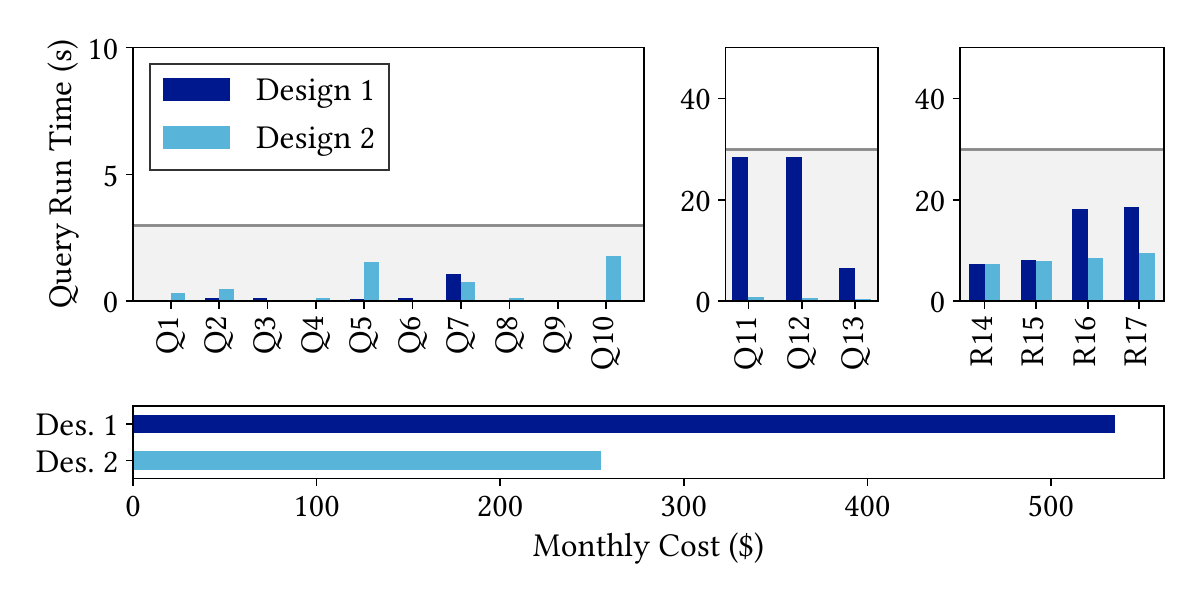}
        \put(49,42.5){\captionlabel{A}}
        \put(68.5,42.5){\captionlabel{B}}
        \put(92,42.5){\captionlabel{C}}
        \put(51.75,10.75){\captionlabel{D}}
    \end{overpic}
    \vspace{-1em}
    \caption{Query performance and operating costs of the same workload on two
    data infrastructure designs.}
    \label{fig:motivation}
    \vspace{-0.5em}
\end{figure}

\section{Conquering the Complex Cloud}
We first illustrate the subtle challenges in cloud infrastructure design and
contrast this experience with using \thesystem{}.

\subsection{When Conventional Wisdom Falls Short}
\label{sec:motivational-examples}
Consider the data processing needs of a movie theater chain.
Under conventional wisdom, they should run an OLTP engine (e.g., Aurora) for
their transactions and a data warehouse (e.g., Redshift) for their analytics.
To show the downsides of this approach, we run a synthetic workload based on a
160~GB version of the IMDB dataset~\cite{leis2015good} comprising transactions,
repeating dashboarding queries \encircle{A} \encircle{B}, and periodic reporting
queries \encircle{C}.
We run Aurora with one db.t4g.medium instance and two Redshift dc2.large nodes
(Design~1).
Figure~\ref{fig:motivation} shows the analytical query latencies.
We aim to keep some queries under 3~s \encircle{A} and others under 30~s
\encircle{B} \encircle{C}.

At first glance, Design~1 appears reasonable. However, consider an alternative
design with just two Aurora db.t4g.medium instances: a primary and replica
(Design~2).
We can run a subset of the queries \encircle{A} \encircle{B} on the Aurora
replica and offload the reporting queries \encircle{C} onto Athena (a serverless
data lake engine).
As shown in Figure~\ref{fig:motivation}, Design~2 saves 2$\times$ on cost
\encircle{D}, meets the performance targets, and even improves query latency on
some queries (up to 48$\times$ \encircle{B} and $2.1\times$ \encircle{C}).
Transaction latency is unaffected on both designs because they run on the
unchanged Aurora db.t4g.medium primary instance.

Design~2 performs better because some queries \encircle{B} 
have predicates on indexed columns, which Aurora can leverage. Redshift
does not support indexes and must use table scans.
The reporting queries \encircle{C} run infrequently, once every four hours, so
they can be offloaded to Athena (a serverless engine) instead of incurring a
high cost on a provisioned but underutilized Redshift cluster.
Athena's serverless burst capability enables the up to 2$\times$ decrease in
query latency \encircle{C}.
These queries cannot meet the performance targets on Aurora; they would run for
over 150 seconds each.

This example shows that an effective design strongly depends on the specifics of
the workload and engines, rather than high-level guiding principles (e.g.,
run transactions on an OLTP engine and analytics on a data warehouse).
Here, the conventional wisdom design is about twice as expensive and an order
of magnitude slower on some queries than Design~2.
An engineer would need an intimate understanding of the engines and workloads to
find such a design, possibly spending a lot of time doing so.
Moreover, because the best design is workload-dependent, it can change in
response to workload shifts, forcing the engineer to redo their work.
These repeated design endeavors are unscalable and difficult to get right,
underscoring the need for a principled and automated alternative.

\subsection{\thesystem{} to the Rescue}\label{sec:brad-system-overview}
With \thesystem{}, users no longer manually design and operate their
infrastructures.
Instead, each virtualized database engine managed by \thesystem{} has a
user-specified design goal (e.g., minimize cost and keep query latency under 30
seconds), and users simply submit their queries and transactions directly to
\thesystem{} as if it were a single engine.
\thesystem{} uses this design goal to optimize the infrastructure to best run
the user's workload---what we call \emph{blueprint planning}.
In this paper, we focus on the models, algorithms, and mechanisms used in
\thesystem{}'s blueprint planner (Section~\ref{sec:brad-key-ideas}).
That said, virtualization comes with many more challenges. In the remainder of
this section, we briefly outline how \thesystem{} tackles them. We leave a more
detailed exploration of this topic for future work.

\sparagraph{Data consistency and freshness.}
Since \thesystem{} can choose to back a (virtual) table with multiple replicas
across engines, consistency and freshness are natural concerns.
In \thesystem{},
\begin{enumerate*}[label=(\roman*)]
  \item a table $T$ will have exactly one ``writer engine'' $E$ (i.e., all DML
    statements affecting $T$ run on the same $E$), and
  \item transactions run on a single engine (Aurora).
\end{enumerate*}
\thesystem{} syncs its table replicas (if any) at a user-defined frequency.
Analytical queries (i.e., read-only queries not part of a transaction) run
against a snapshot, but the snapshot can be stale up to the last sync.
All transactions always run on the latest snapshot. 
This approach provides similar freshness to existing
solutions~\cite{karagiannis2013scheduling}.

\sparagraph{Transformations.}
Modern data infrastructure designs typically use ELTs~\cite{etl-vs-elt,zeroetl},
meaning that tables are first replicated into a data warehouse and then
transformed inside the warehouse using DML statements.
\thesystem{} supports this model, as it already syncs table replicas across
engines; these transformations would thus run as regular DML statements that
modify the logical tables in \thesystem{}.
Running these transformations on a schedule is orthogonal to \thesystem{} and
can be handled using an external tool.

\sparagraph{SQL dialects and semantics.}
Different database engines can have different SQL dialects and semantics,
meaning the same SQL statement may not be executable on every engine. Currently,
we assume that \thesystem{} can detect the subset of its engines that can
correctly run a given SQL query; \thesystem{} will ensure it only routes the
query to those eligible engines (see Section~\ref{sec:brad-query-routing}). SQL
dialect translation techniques~\cite{sqltranslation} may enable \thesystem{} to
expand a query's set of eligible engines; we leave this to future work.

\begin{figure}[t]
  \includegraphics[width=0.9\columnwidth]{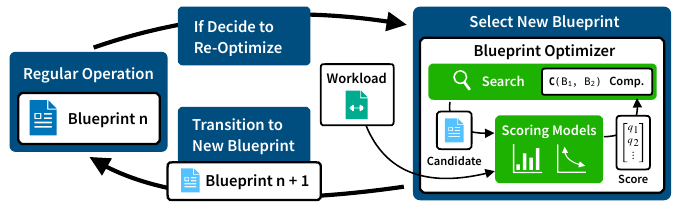}
  \vspace{-0.25em}
  \caption{\thesystem{}'s blueprint planning life cycle.}
  \label{fig:bpp-lifecycle}
\end{figure}

\section{\thesystem{}'s Blueprint Planner: Key Ideas}\label{sec:brad-key-ideas}
We now describe the key ideas behind \thesystem{}'s blueprint planner. We
continue with further implementation details in Section~\ref{sec:brad-details}.

\subsection{The Blueprint Planning Life Cycle}\label{sec:bp-lifecycle}
\thesystem{} automatically designs and operates data infrastructures using a
blueprint planning \emph{life cycle}, which we depict in
Figure~\ref{fig:bpp-lifecycle}.
The core idea is to select the ``best'' blueprint for the user's workload,
operate the infrastructure according to the blueprint, and then trigger
re-optimization if the workload changes%
\ifdefined\BradExtended
~(Section~\ref{sec:brad-triggers}).
\else
.
\fi
Concretely in \thesystem{}, blueprints are infrastructure plans that contain
\begin{itemize}[leftmargin=*]
  \item The engines to include in the underlying data infrastructure.
  \item The provisioning configuration to use for each engine when applicable
  (e.g., instance type, the number of nodes to use).
  \item The placement of data tables and replicas on the engines.
  \item A policy for routing queries to the engines in the infrastructure.
\end{itemize}
Below, $B$ is an example blueprint describing an infrastructure comprising
Aurora (provisioned with one db.r6g.xlarge instance) and Athena. Table $T_1$ is
placed on Aurora, and $T_2$ is replicated on Aurora and Athena.
The routing policy consists of concrete query assignments chosen during
blueprint optimization (query $q_1$ to Aurora and $q_2$ to Athena) (see
Section~\ref{sec:search-key-idea}) and an online policy $P(q)$ that selects an
engine for a given query $q$ (see Section~\ref{sec:routing-key-idea}).

{
  \footnotesize
  \vspace{-1em}
  \[
    B =
    \begin{cases}
      \{\mathrm{Aurora}, \mathrm{Athena}\} & \mathbf{Engines} \\
      \{(\mathrm{Aurora}, \mathrm{db.r6g.xlarge}, 1)\} & \mathbf{Provisioning} \\
      \{T_1 \rightarrow \mathrm{Aurora}, T_2 \rightarrow \mathrm{Aurora}, T_2 \rightarrow \mathrm{Athena}\} & \mathbf{Placement} \\
      \{q_1 \rightarrow \mathrm{Aurora}, q_2 \rightarrow \mathrm{Athena}, P(q) \} & \mathbf{Routing}
    \end{cases}
  \]
}

To automatically find an optimized blueprint, \thesystem{} needs a mechanism
to
\begin{enumerate*}[label=(\roman*)]
    \item quantify the utility of (i.e., assign a ``score'' to) candidate
    blueprints on the user's workload (Section~\ref{sec:scoring-key-idea}), and
    \item to systematically search over the blueprint design space
    (Section~\ref{sec:search-key-idea}).
\end{enumerate*}
Here, a workload is a representative (but not necessarily exhaustive) list of
expected queries and DML statements along with dataset statistics (e.g., its
size).
Concretely, \thesystem{} obtains a user's workload by logging their transactions
and queries (see Section~\ref{sec:realizing-lifecycle}).

Scoring a blueprint is challenging because multiple factors influence a
blueprint's utility (e.g., performance, cost), and different users may have
different design goals (e.g., maximizing performance vs. minimizing cost).
Consequently, \thesystem{} assigns \emph{vector scores} to its candidates, which
comprise three components:
\begin{enumerate}[leftmargin=*]
  \item \textbf{Workload Performance.} \thesystem{} predicts the run time of the
    queries and transactions in the workload on the blueprint.
  \item \textbf{Operating Cost.} The monetary cost of operating the data
    infrastructure and routing policy specified by the blueprint.
  \item \textbf{Transitions.} The time and monetary cost of
    \emph{transitioning} the underlying infrastructure to the candidate blueprint.
\end{enumerate}
\ifdefined\BradExtended
For example, a vector score would look like $\left[ q_1 \; q_2 \; \dots \; c \;
t_T \; c_T \right]^\intercal$ where the $q_i$s are predicted query and
transaction latencies, $c$ is the operating cost, and $t_T$ and $c_T$ are the
transition time and cost.
\fi
\thesystem{} uses a set of learned models to assign values to all three
components, which we discuss next in Section~\ref{sec:scoring-key-idea}.

Users express their ``design goals'' to \thesystem{} by providing a comparator
function that ranks the vector scores (Section~\ref{sec:brad-bp-comparator}),
analogous to the comparators used in sort routines~\cite{comparator}. 
For example, one such goal could be to design an infrastructure that minimizes
cost while maintaining a performance constraint (e.g., a latency SLO, see
Section~\ref{sec:brad-bp-comparator}). The comparator would therefore rank
blueprints by their operating cost while treating blueprints that are predicted
to not meet the latency constraint as having an infinite cost.

\subsection{Blueprint Scoring}\label{sec:scoring-key-idea}
In the blueprint's score vector, the main challenge is
predicting the performance of the workload on the blueprint. For \thesystem{},
this means estimating the latencies of the queries in the workload on each of
\thesystem{}'s engines while taking into account their provisioning and load. We
discuss scoring in more depth in
Section~\ref{sec:brad-bp-scoring-details}.

\subsubsection{Query Run Times}
Prior work has proposed run time prediction methods for use in
query optimization~\cite{SunL19}, workload
scheduling~\cite{WagnerK021}, resource management~\cite{AutoWLM}, and
maintaining SLOs~\cite{ChiMHT11}. These methods require the query's physical
execution plan as input. For example, DBMS cost models and traditional
predictors use hand-derived heuristics to understand the cost of each physical
operator~\cite{DugganPCU14, WuCZTHN13, AkdereCRUZ12, LiKNC12}. Advanced methods
featurize the physical query plans and train deep learning models to predict
their run time~\cite{SunL19, MarcusP19, HilprechtB22, WuYYZHLLZZ22,
marcus2019neo, marcus2022bao}.
\thesystem{} cannot directly use these methods because it cannot always get a
physical plan. For example, \thesystem{} may need to predict the run time of a
query on an engine that is not running or does not have the relevant data loaded
(e.g., to decide whether to start Redshift and/or move a table there).
\thesystem{} must also account for the effects of provisioning and
system load.

\thesystem{} addresses these challenges using a graph neural network (GNN) and
two analytical models.
We design a new GNN that predicts a query's run time using the query's SQL as
input (i.e., relies only on logical features) for an unloaded engine on a fixed
provisioning.
Our GNN's novelty is that it featurizes a query based on its SQL text and
data properties.
In contrast, existing models featurize queries using their physical query plans.
We then use two analytical models, based on Amdahl's law~\cite{amdahlslaw} and
queuing theory~\cite{queuing-book-mhb-13} to adjust this model's estimates for
different provisionings and system loads, respectively.
We take this approach because making such predictions with a single
model is expensive and hard to realize due to the need for diverse run
time observations across various query types, provisionings, and system loads.
We describe the details of our analytical models in
Section~\ref{sec:brad-bp-scoring-prov}; they provide an acceptable accuracy and
enable \thesystem{} to find effective blueprints
(Section~\ref{sec:eval-e2e-scenarios}).

\sparagraph{GNN model and query featurization.}
We use a GNN model with a novel query featurization that depends only
on logical query properties (e.g., the join template, join/filter selectivity)
and dataset statistics (e.g., estimated join selectivity).
Our design is based on the key observation that most query optimizers are
deterministic: they will choose similar query plans with similar run times for
queries with similar features and statistics. Thus, we identify these features
and then model them with a novel graph structure.
As a result, even without physical plans, our model learns the optimizer’s
behavior and makes accurate predictions for queries similar to the training
queries in our featurization space (Section~\ref{sec:eval-accuracy}).
We use the same approach to predict the amount of data a query scans (to
estimate Athena's query cost).
We describe the featurization and graph structure in more detail in
Section~\ref{sec:brad-bp-scoring-rt}.

\subsubsection{Model Bootstrapping}\label{sec:brad-bp-bootstrap}
\thesystem{} is designed to be gradually deployed onto an existing
infrastructure running one or more of our component engines.
When first deployed, \thesystem{} observes the running workload and gathers
performance data (e.g., query run times) for each engine in a brief
``bootstrapping phase.''
\thesystem{} then uses this data to train these aforementioned models.
Once complete, \thesystem{} then begins to actively optimize the infrastructure
using its blueprint planner.
Avoiding a bootstrapping phase would require having performance models that are
fully transferable across workloads and datasets, which we leave to future
work.

\subsection{Blueprint Search}\label{sec:search-key-idea}
Exhaustively searching the entire design space is intractable for most workloads
because it is exponentially large with respect to the number of tables and
queries.
Given this challenge, \thesystem{} must use an efficient search algorithm that
finds optimized blueprints without visiting the entire search space.

\thesystem{} uses a greedy beam-based search~\cite{lowerre1976harpy} over the
blueprint's routing policy (i.e., query-to-engine mapping), which directly
impacts the workload's performance.
Blueprints are optimized for a workload which contains a representative list of
expected queries.
The idea is to incrementally expand a set of top-$k$ blueprints (the ``beam'')
by examining queries in the workload one-by-one.
For each blueprint in the current top-$k$, the planner takes the next query and
assigns it to each of the three engines, creating three new candidate
blueprints.
It then places tables according to these routing decisions (e.g., if a query
accessing table $T$ is routed to $E$, then a copy of $T$ is placed on $E$).
After each step, the planner only keeps the top-$k$ blueprints, and it repeats
until all the queries have been assigned.
\thesystem{} runs this beam search for each provisioning ``near'' the current
one (in computational resources) and returns the best-scoring blueprint.
\thesystem{} uses a beam of size 100 (i.e., $k = 100$). We analyze and discuss
our search algorithm in more detail in Section~\ref{sec:bp-search-details}.

\subsection{Operating Blueprints: Query Routing}\label{sec:routing-key-idea}
Once \thesystem{} has chosen a blueprint, the final step is to use it to serve
the workload.
To route queries, \thesystem{} first consults the query-to-engine assignment in
its blueprint. If the query is in the assignment, \thesystem{} uses this pre-planned
routing decision.
Otherwise, it uses an online routing policy.
\ifdefined\BradExtended
The key challenge in designing this policy is that it must make intelligent
routing decisions without imposing an undue overhead (i.e., doing so within tens
of milliseconds). This efficiency constraint precludes the use of
computationally expensive models, such as the query run time model we use for
blueprint scoring.
To address this challenge, we make two key observations.
First, in tasks like run time prediction, prior work showed that classification
is easier than regression in terms of model efficiency and
accuracy~\cite{ZhuCDCPWZ23, preferencelearning}.
Thus, we can cast this routing problem as a classification problem and leverage
a lightweight classifier.
Second, this online routing policy can be trained during blueprint planning,
which runs off of the critical path.
This workflow gives us the opportunity to bootstrap the online routing policy
using a more sophisticated but computationally expensive model, e.g., our query
run time model.

We leverage these observations to design \thesystem{}'s online routing policy.
\fi
\thesystem{}'s online policy is a decision forest that, for a given query,
produces a ranking of the engines (most preferred routing to least).
\thesystem{} routes the query to the highest-ranked engine that has all the
tables the query accesses.
As input, the forest takes the estimated scan cardinality of each table that the
query accesses; these cardinalities can be computed using an off-the-shelf
cardinality estimator.
The forest is trained as the final step in blueprint optimization using run
times that our query run time model predicts (the engine with the lowest
predicted run time is the most preferred).
Inference over the decision forest is fast and does not impose an undue overhead
on the queries.
\ifdefined\BradExtended
We empirically evaluate the effectiveness and overhead of our routing policy in
Section~\ref{sec:eval-routing}.
\fi
We discuss additional practical details for query routing in
Section~\ref{sec:brad-query-routing}.

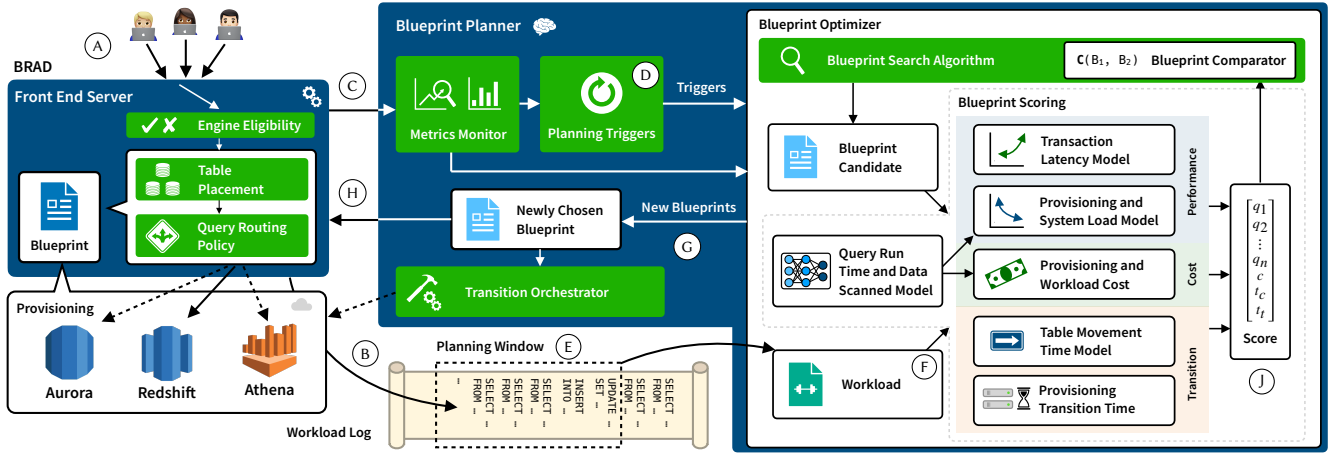
\begin{figure*}
  \centering
  \begin{overpic}[width=\textwidth]{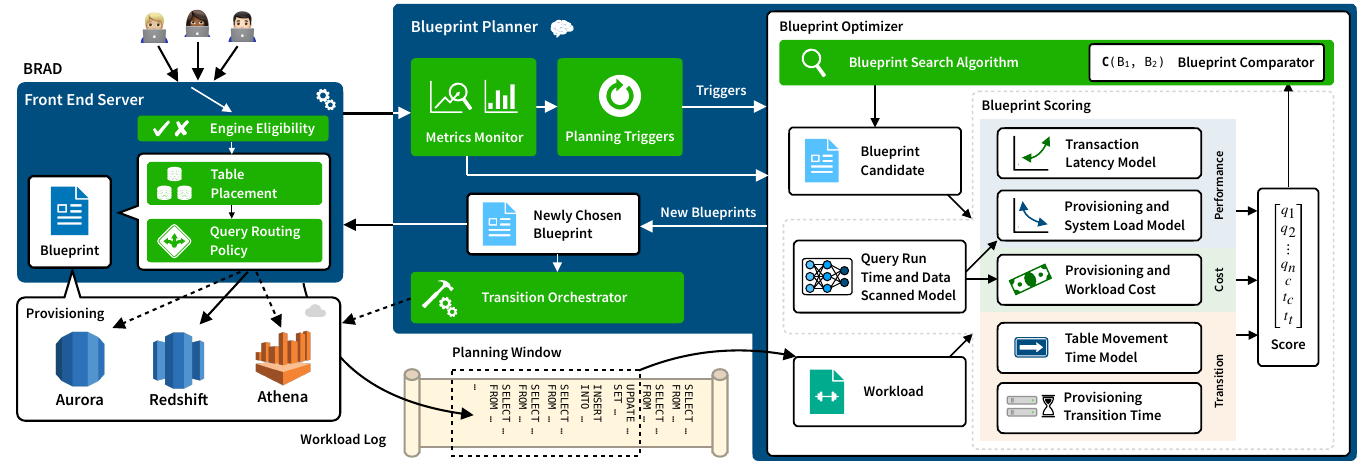}
    \put(7,30){\captionlabel{A}} 
    \put(27,7){\captionlabel{B}} 
    \put(26,27){\captionlabel{C}} 
    \put(47.9,27.8){\captionlabel{D}} 
    \put(42.2,7.5){\captionlabel{E}} 
    \put(68.8,6){\captionlabel{F}} 
    \put(51,15){\captionlabel{G}} 
    \put(26,19){\captionlabel{H}} 
    \put(94,5){\captionlabel{J}} 
  \end{overpic}
  \vspace{-1em}
  \caption{A detailed view of \thesystem{}'s architecture and its end-to-end
    blueprint planning life cycle.}
  \label{fig:brad-bpp}
\end{figure*}

\section{\thesystem{}'s Blueprint Planner: Details}\label{sec:brad-details}
In this section we outline the implementation details behind \thesystem{} and
provide additional details about \thesystem{}'s blueprint planner.

\subsection{Realizing the Blueprint Planning Life Cycle}\label{sec:realizing-lifecycle}
Figure~\ref{fig:brad-bpp} depicts \thesystem{}'s system architecture, which
implements the blueprint planning life cycle using two components:
\begin{enumerate*}[label=(\roman*)]
    \item a front-end server responsible for interfacing with clients and
    operating the blueprint, and
    \item a blueprint planner that monitors the workload and selects new
    blueprints when appropriate.
\end{enumerate*}
We describe the architecture by walking through the blueprint planning
life cycle.

The life cycle begins at \thesystem{}'s front-end server
(Figure~\ref{fig:brad-bpp}~\encircle{A}).
Users submit SQL queries to the server, which get routed to a suitable engine
for execution (Section~\ref{sec:routing-key-idea}).
Crucially, to keep track of the executing workload, the front end
\begin{enumerate*}[label=(\roman*)]
  \item logs the queries it receives \encircle{B}, and
  \item collects metrics about the workload (e.g., transaction latency, query
    latency).
\end{enumerate*}
The blueprint planner monitors these metrics \encircle{C}, alongside others it
retrieves from the underlying engines (e.g., CPU utilization) and triggers
blueprint optimization when they exceed or fall below specified thresholds
\encircle{D}%
\ifdefined\BradExtended
~(Section~\ref{sec:brad-triggers})
\else
.
\fi

When starting blueprint optimization, \thesystem{} first extracts the queries
that ran during its planning window (a sliding window of a configurable length
\encircle{E}) from its workload log.
These queries, along with dataset statistics (e.g., the sizes of the tables),
are passed to the optimizer and represent the workload that \thesystem{} uses
when scoring candidate blueprints \encircle{F}.
\thesystem{}'s optimizer then searches
over valid blueprints (Section~\ref{sec:search-key-idea}), scores them
(Section~\ref{sec:brad-bp-scoring-details}), and returns the best-scoring
blueprint \encircle{G} (Section~\ref{sec:brad-bp-comparator}).
The planner then transitions the infrastructure to the chosen blueprint and
passes it to the front end \encircle{H}.
The front end uses this blueprint until the next one is chosen, completing the
blueprint planning life cycle.

\subsection{Additional Blueprint Scoring Details}\label{sec:brad-bp-scoring-details}

\subsubsection{Query Run Time and Data Scanned}\label{sec:brad-bp-scoring-rt}
As discussed in Section~\ref{sec:scoring-key-idea}, \thesystem{} uses a graph
neural network with a novel query featurization to predict a query's run time
and the amount of data it scans. We now describe the featurization and model in
more detail.

\begin{figure}
  \includegraphics[width=\columnwidth]{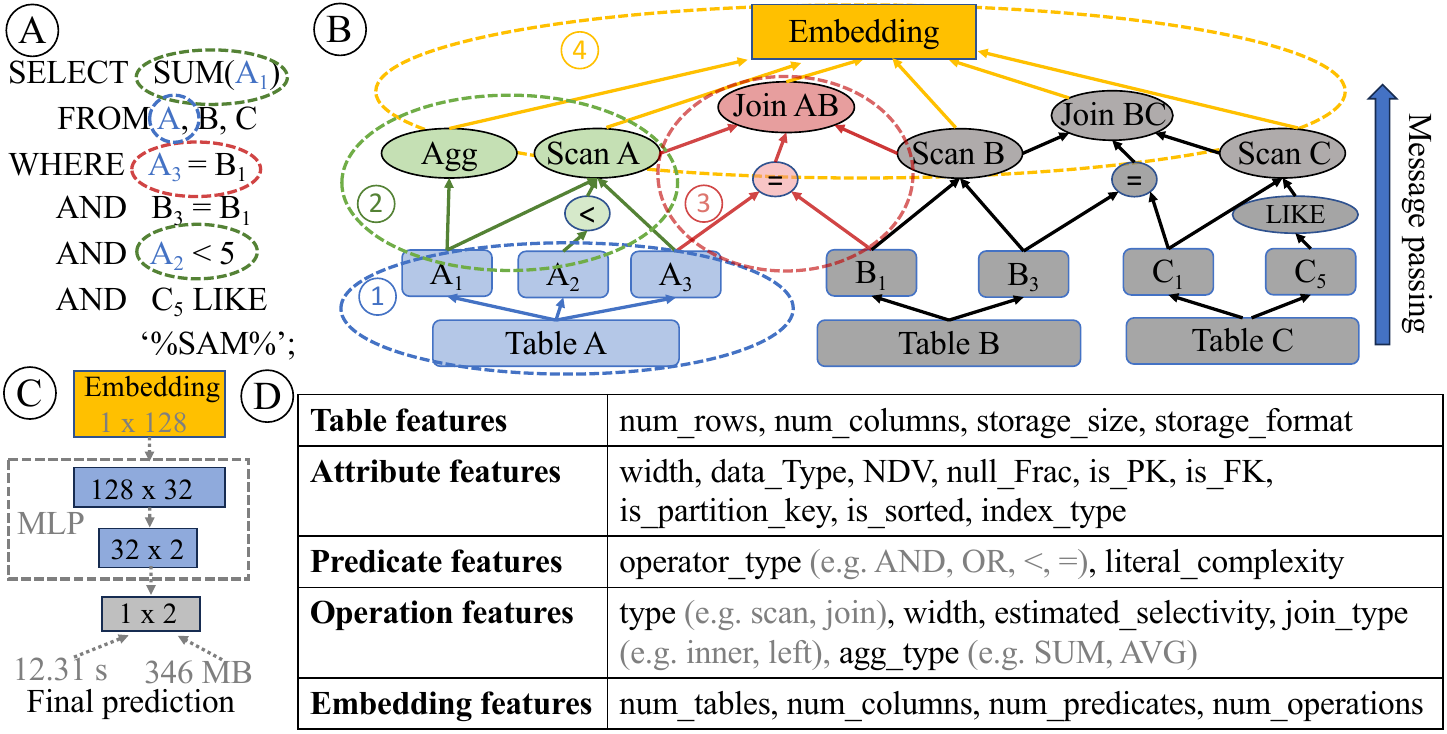}
  \vspace{-1.75em}
  \caption{The query featurization used by our model, which predicts a query's run
    time and the amount of data it scans.}
  \label{fig:cost_model_query_feature}
  \vspace{-1em}
\end{figure}

\sparagraph{Query featurization.} 
As discussed, existing run time predictors require the query’s physical
execution plan as input, which is not always available in \thesystem{}'s setting
(Section~\ref{sec:scoring-key-idea}).
Thus, we design a new graph featurization approach to encode information, such
as filters, joins, and group-bys, purely from a query's SQL.
This logical featurization differentiates our GNN from existing
models~\cite{SunL19, MarcusP19, HilprechtB22, WuYYZHLLZZ22, marcus2019neo,
marcus2022bao}.
Figure~\ref{fig:cost_model_query_feature} shows an example procedure to extract
such a query feature graph.
It has five types of nodes, each with distinct node features (shown in Table
\encircle{D} in Figure~\ref{fig:cost_model_query_feature}) and edges
representing the dependencies between these nodes.
We parse the query's SQL to extract the tables, columns, predicates, and logical
operations it involves and construct the feature graph in four steps.

First, we extract the tables and columns involved in the query. 
We show an example table $A$ and its relevant columns $A_1$, $A_2$, $A_3$ in
blue in Figure~\ref{fig:cost_model_query_feature}~\encircle{A}.
We create a table node for each table and a column node connecting to a table
node for each column (Figure~\ref{fig:cost_model_query_feature}~\encircle{B}).
Second, we extract the operations on a single table, such as ``scan'',
``aggregate'', and ``group-by'', as highlighted in green in
Figures~\ref{fig:cost_model_query_feature}~\encircle{A} and
\ref{fig:cost_model_query_feature}~\encircle{B}. 
Specifically, we create a predicate node for each filter predicate and connect
the column nodes involved in this predicate to it. 
We extract the features of the predicate nodes using an approach similar to
recent work~\cite{SunL19, HilprechtB22}. 
We connect these predicate nodes to their parent operation node.
For operations without a predicate, such as ``aggregate'', we just connect them
to the relevant column nodes. 
Third, we parse the operations involving multiple tables, e.g., ``join'', as
highlighted in red.
The children of each join operation node are the relevant scan operation nodes
and the join predicate node extracted the same way as a filter predicate.
We list the features for all node types in
Figure~\ref{fig:cost_model_query_feature}~\encircle{D}.
Finally, as highlighted in yellow in
Figure~\ref{fig:cost_model_query_feature}~\encircle{B}, all the operation nodes
are connected to the embedding node, which aggregates the overall information.

This generic query graph representation does not encode any information about a
physical plan (e.g., join order or physical operators).
The features we use can all be independently derived without relying on
\thesystem{}'s underlying engines.
We use a value of $-1$ for the aforementioned node features if the feature is
unavailable.

\sparagraph{Selectivity estimates.}
Our model also uses an estimate of the selectivity of each operation node. This
is because the selectivity influences the chosen physical plan, and thus the
query's run time as well.
\thesystem{} collects simple statistics about each table (e.g., histograms)
using analysis pass over the underlying data. \thesystem{} then uses the
Selinger method~\cite{selingeropt-selinger79} to make these estimates because of
its simplicity and negligible overhead.
The choice of estimator is orthogonal to our model; other methods are also
applicable~\cite{factorjoin-wu23, robust-cardinality-negi23}.

\sparagraph{Graph neural network.} 
Inspired by the zero-shot run time predictor~\cite{HilprechtB22}, we use a graph
neural network to model interactions between nodes and propagate information
through the feature graph.
We construct one multi-layer perception (MLP) node encoder for each level that
embeds the node features into a fixed-length vector.
Then, we create another MLP for message passing through
edges~\cite{GilmerSRVD17}.
At each internal node, we sum its children's embedded vectors, concatenate them
with its own vector, and feed the resulting vector to the MLP to get a new
vector.
The message passing stops at the last level when the ``embedding node'' has
aggregated all of the query's information into a single vector.
This vector is fed to two MLPs
(Figure~\ref{fig:cost_model_query_feature}~\encircle{C}) to predict the query's
run time and amount of data scanned.
The entire model is trained end-to-end.

\sparagraph{Discussion.}
By using logical features and selectivity estimates, our model implicitly learns
a query optimizer’s behavior and makes accurate predictions for queries similar
to our training queries (see Section~\ref{sec:eval-accuracy}).
Our model may not work well in some cases where the testing queries are
\emph{significantly} different from our training queries. For example, suppose
the model was only trained on short running queries joining small tables and the
user then submits a long running query joining large tables.
Then, the engine could possibly choose a join operator/order that it has never
chosen before, and our model (and possibly all existing run time models) would
be unlikely to correctly predict that query's run time.
However, recent work shows that in practical workload traces, a large portion of
queries are repeating and ``similar'' to prior queries~\cite{wu2024stage}.
Thus, encountering a query that greatly impacts the selected plan
in a way that was not captured in the training data should be rare.

\subsubsection{Provisioning and System Load}\label{sec:brad-bp-scoring-prov}
Next, we describe the two analytical models we use to adjust our GNN's run time
estimates for different provisionings and system load.

A query's run time consists of two components: the time spent running and the
time it queues due to system load.
Thus, \thesystem{} models a query's complete run time $R$ as $R(G, \rho) = P(G)
+ W(\rho)$, where $G$ is the run time given by our GNN, $P(\cdot)$ is a model
that accounts for the engine's provisioning, and $W(\cdot)$ is the time spent
queuing depending on the system's utilization (load) $\rho$.

\sparagraph{Compute resources.}
\thesystem{} uses $P(G) = \left( c_1 (b/d) + c_2 \right) G$, which we derive
from Amdahl's law~\cite{amdahlslaw}.
We model the query's run time as having two parts: (i) one that will decrease
(or increase) if the engine's provisioning is changed to have more (or less)
compute resources, and (ii) a fixed part that will not change even with more
resources.
Here $c_1 G$ represents part (i) and is multiplied by $b/d$, which is the ratio
between the resources available on the ``base'' and ``destination''
provisionings.
The base is the provisioning on which the graph neural network was trained. The
destination is the engine's provisioning in the candidate blueprint. 
The $c_2 G$ term is part (ii).
\thesystem{} uses the total number of vCPUs in the provisioning to represent the
available compute.
We learn one set of constants $c_1$ and $c_2$ for each engine (Aurora and
Redshift) using least squares linear regression~\cite{bishop2006pattern} on the
query workload.
This approach assumes that provisioning changes do not cause significant query
plan changes that would affect the run time.
Empirically, we find this model to be sufficient for blueprint planning
(Section~\ref{sec:eval-e2e-scenarios}).

\sparagraph{System load.}
\thesystem{} models $W(\rho_r)$ using queuing theory, approximating an engine as
an M/M/1 system~\cite{queuing-book-mhb-13}. We use $W(\rho_r) = -K
\left( 1 - \rho_r \right)^{-1} \log\left(\rho_r^{-1} (1-q) \right)$ which models
the $q$-th percentile queuing time on a system with utilization
$\rho_r$~\cite{queuing-systems-15}. $K$ represents the mean processing time,
which we estimate as the average $P(G)$ for all queries assigned to the engine.
We approximate an engine's utilization using its measured CPU utilization. In
our experiments, \thesystem{} optimizes for a p90 latency constraint, so we use
$q = 0.9$.
We use this model as it provides a simple closed-form expression for wait time.
Not all engines are M/M/1 systems; for example, the query arrival distribution
may not be exponential and/or the engine may process queries in parallel.
However, we empirically find that this simple model is sufficient for blueprint
planning (Section~\ref{sec:eval-e2e-scenarios}).

\sparagraph{Adjusting $\rho_r$.}
$W(\rho_r)$ relies on a representative $\rho_r$.
\thesystem{} cannot directly use CPU utilization because the candidate blueprint
may use a different query routing as the current blueprint, thereby imposing
different loads.
Instead, we assume that a query's ``load'' is proportional to its run time.
\thesystem{} thus scales its observed CPU utilization by a factor: the sum of
the run times of the queries routed to the engine in the candidate blueprint
divided by the sum of the run times that actually ran on the engine in the last
planning window.
\ifdefined\BradExtended
If no queries ran on the engine in the previous planning window (e.g., the
engine was paused), \thesystem{} scales the query's predicted run times to a
load value using a learned proportionality constant.
\fi

\subsubsection{Transaction Latency}\label{sec:brad-bp-scoring-txns}
\thesystem{} estimates transactional latency on a candidate blueprint to
ensure it provisions Aurora appropriately for the transactional load it
experiences.
In general, estimating a transaction's run time is a hard problem due to the
many factors that can influence its latency (e.g., lock contention, buffer pool
state, etc.)~\cite{oltp-mozafari13}.
We make a simplifying assumption that the transactional workload is uncontended
and consists of TPC-C-like~\cite{tpcc} indexed point reads and writes made
interactively over the network.
For this setting, we use an analytical function that models a transaction's
latency as a function of system utilization: $R(\rho_t) = a/(M - \rho_t) + b$.
Here, $R(\rho_t)$ is the overall transactional latency. $a$, $b$, and $M$ are
workload-specific learned constants. $\rho_t \in [0, 1]$ is the system
utilization; we require that $M > \rho_t$. We use CPU utilization as a simple
proxy metric for $\rho_t$.
This model captures that transaction latency increases rapidly as $\rho_t$
approaches $M$, like it would on an overloaded
system~\cite{queuing-book-mhb-13}.
We derive this model empirically; it is loosely based on the queuing theory
equations for an M/M/1 system~\cite{queuing-book-mhb-13}.

\sparagraph{Adjusting $\mathbf{\rho_t}$.}
The candidate blueprint's $\rho_t$ may not be the same as the measured $\rho_t$
on the current blueprint (e.g., due to a changed provisioning and/or query
routing).
\thesystem{} applies two scaling factors to compensate.
First, it multiplies $\rho_t$ by the ratio of vCPUs between the candidate
blueprint's provisioning and the current blueprint's provisioning.
Second, it applies the same query load scaling factor mentioned in
Section~\ref{sec:brad-bp-scoring-prov} to account for query movement.

\subsubsection{Operating Cost and Transitions}\label{sec:brad-bp-scoring-rest}
A blueprint's cost comprises provisioning costs, Athena query costs, and storage
costs. For Aurora and Redshift, \thesystem{} uses AWS' on-demand instance
pricing~\cite{aurora-pricing,redshift-pricing}.
Currently \thesystem{} only considers Aurora I/O optimized instances, which do
not bill I/O usage~\cite{aurora-ioopt}.
Since Athena bills by the amount of data scanned, \thesystem{} uses its
data scanned predictions (Section~\ref{sec:brad-bp-scoring-rt}) along with
Athena's scan pricing~\cite{athena-pricing}.
For storage costs, \thesystem{} models a table's size as $k|T|$ where $|T|$ is
the number of rows in the table and $k$ is an engine and table-specific
constant.
To compare blueprints, \thesystem{} normalizes costs to be in \$/hour.

\sparagraph{Table movement.}
\thesystem{} currently moves tables via S3 (i.e., export to S3 and import from
S3).
It estimates the movement time as $S/k_e + S/k_i$, where $S$ is the physical
size of the table and $k_e$ and $k_i$ are empirically measured export and import
rates.
These rates are engine-specific but independent of the engine
provisioning, which we confirmed empirically.
AWS does not charge for S3 transfers between AWS services, so \thesystem{} does
not incur movement costs.

\sparagraph{Aurora provisioning time.}
\thesystem{} estimates Aurora's provisioning time as the number of instance
changes multiplied by a fixed amount of time (we empirically measured 5
minutes).
Removing replicas is considered to take no time since \thesystem{} does not need
to wait for the completion of removal to start using the next blueprint.

\sparagraph{Redshift provisioning time.}
The time it takes to complete a Redshift provisioning change depends on whether
it can be done using an elastic resize or not~\cite{redshift-resize}.
For elastic resizes, \thesystem{} uses AWS' published estimate of 15
minutes~\cite{redshift-resize}.
\thesystem{} estimates classic resizes to take
$|R|/k$ where $|R|$ is the physical size of the data in the Redshift cluster.
We empirically observed $k$ to be approximately 18 megabytes per second.
Pausing Redshift is also considered to take no time for the same reason as
Aurora replica removals.

\subsection{Additional Query Routing Considerations}\label{sec:brad-query-routing}
Upon receiving a query, \thesystem{} selects a suitable engine in two steps:
\begin{enumerate*}[label=(\roman*)]
  \item determine the set of engines that are \emph{able} to execute the query,
    and then
  \item select the most suitable (e.g., best performing) engine from this set.
\end{enumerate*}
In this section, we describe step (i). For (ii), \thesystem{} uses the online
routing policy described in Section~\ref{sec:routing-key-idea}. 

In \thesystem{}, an engine's eligibility to run a query depends on the
\emph{table placement} and its \emph{functionality support}.
Table placement is the set of engines that hold a copy of a table and is
governed by the blueprint; \thesystem{} currently only routes a query to an
engine if it has a copy of every table the query accesses.
\thesystem{} parses the SQL query to extract the tables it references and
compares them against the blueprint's table placement.
During blueprint planning, \thesystem{} ensures that all tables are placed
together on at least one engine to so that it can always run a query that joins
any subset of tables.

A query may also use specialized functionality only available on a subset of the
engines (e.g., vector similarity search~\cite{pgvector,rag2020}).
By taking functionality into account, \thesystem{} ensures that it only selects
engines that can run the query.
Automatically determining the ``specialized functionality'' that a query uses is
something we leave to future work.  \thesystem{} currently uses keyword matching
against pre-specified keyword lists to determine if a query uses such
functionality (e.g., the presence of the \texttt{<=>} operator would imply that
the query uses vector similarity search).

\subsection{Additional Blueprint Search Details}\label{sec:bp-search-details}
\ifdefined\BradExtended
Algorithm~\ref{alg:brad-bp-search} contains the pseudocode for \thesystem{}'s
greedy beam blueprint search algorithm, which we outline in
Section~\ref{sec:search-key-idea}.
\fi
The intuition for using a top-$k$ beam search instead of a na\"ive greedy search
lies in the nature of the search space.
At the beginning of the search, assigning queries to some engines may be better
(e.g., prefer Athena, which is pay-per-query, instead of Redshift where you pay
for provisioning).
But after assigning more queries, this trade-off changes (e.g., there are enough
queries to justify running Redshift).
Keeping a set of promising candidates helps \thesystem{} balance these changing
trade-offs. 
We search over nearby provisionings because \thesystem{} handles
gradual workload changes; the next best provisioning is likely to be near the
current provisioning.

\sparagraph{Discussion.}
Beam search works well empirically in our setting for two reasons.
First, \thesystem{} uses a large beam ($k = 100$), which helps
prevent some promising candidates from being eliminated too early.
Second, the queries in our workload have a skewed arrival frequency (i.e., some
queries arrive more frequently than others).
This property was also observed by our industrial partners in their real-world
workloads~\cite{wu2024stage}.
As a result, \thesystem{} processes queries in decreasing order of arrival
frequency. Along with using a large beam, this processing strategy makes it more
likely for ``important'' (i.e., frequently occurring) queries to be assigned to
the best engine.

\sparagraph{Analysis.}
Let $m$ be the number of engines considered, $q$ be the number of queries in the
workload, and $p$ be the number of distinct provisionings considered.
This algorithm considers $O(kmqp)$ candidate blueprints. Currently, \thesystem{}
has $m = 3$ and uses $k = 100$.
\ifdefined\BradExtended
We evaluate our algorithm empirically in Section~\ref{sec:eval-bp-planner}.
\fi

\subsection{Blueprint Comparator: Minimizing Cost}\label{sec:brad-bp-comparator}
As discussed in Section~\ref{sec:bp-lifecycle}, end-users need to define a
comparator function, which imposes an ordering on blueprint vector scores. This
comparator is how users convey their infrastructure design goals to
\thesystem{}. A common goal is to minimize an infrastructure's operating
costs while maintaining a service level objective (SLO) (e.g., p90 query latency
should be under 30 seconds). We use this design goal when evaluating
\thesystem{} in Section~\ref{sec:evaluation}. We now describe how this goal is
implemented as a comparator.

Given two blueprints ($B_1$, $B_2$), the general idea is to map their vector
scores to scalar costs ($W_1$, $W_2$); the candidate with the lower cost is
considered better.
A simple mapping would be to use the blueprint's operating cost when the predicted
query latency falls under the desired latency constraint and an infinite cost
otherwise (to indicate infeasibility). However, this mapping does not
consider the time and cost of transitioning to the candidate.
Instead, our approach is to weigh the cost of operating each blueprint using the
transition time $T_T$ and a user-defined ``benefit period'' $T_B$. This period
represents how long the user expects the workload to ``benefit'' from the new
blueprint. Concretely, we use
\begin{align*}
  W = P^\gamma C_0 T_T + C_T + C T_B \qquad
  P = 1 + \mathrm{max}\left( t / t_{\text{SLO}}, q/q_{\text{SLO}} \right)
\end{align*}
$C_0$ represents the current blueprint's operating cost, $C$ is the candidate
blueprint's predicted operating cost and $C_T$ is the transition cost.
$P \in [1, \infty)$ is a penalty multiplier that grows as the
current blueprint approaches and exceeds the performance constraints (e.g.,
because the workload changes). $\gamma$ is a user-chosen weight (we use $\gamma
= 2$).
$t$ and $q$ represent the transaction and analytical latency measured on the
current blueprint; $t_{\text{SLO}}$ and $q_{\text{SLO}}$ represent the
user-specified performance constraints for these values.
Users can declare multiple such constraints (e.g., for different queries).

When the current blueprint exceeds the performance constraints, the first term
in the equation on the left will dominate. Thus \thesystem{} will prefer
candidate blueprints that are faster to transition to. Otherwise, \thesystem{}
weighs the operating costs by the time spent transitioning versus running the
new blueprint.
This means \thesystem{} will still consider blueprints requiring
very expensive transitions (high $T_T$) but will only select them if their
benefit is large enough (low $C$ during $T_B$).
If a candidate blueprint has a predicted transactional or analytical latency
greater than $t_{\text{SLO}}$ or $q_{\text{SLO}}$, we just assign an infinite
cost. If all of the candidates are infeasible, \thesystem{} will ask the user to
change their constraints.

\ifdefined\BradExtended
\begin{algorithm}[t]
  \caption{\thesystem{}'s greedy beam blueprint search algorithm.}\label{alg:brad-bp-search}
  \begin{algorithmic}
    \Require $B_0$: Current blueprint, $\;W$: Workload, \\$\;\mathbf{Score}(\cdot, \cdot, \cdot)$: Blueprint scoring function
    \Ensure $B^*$: Best found blueprint
    \State
    \Function{DoSearch}{$P$: Provisioning}
      \State Sort queries in $W$ in decreasing order of arrival probability and
        then largest predicted speedup across engines
      \LComment{Initial blueprint with provisioning $P$ and no routed queries}
      \State $T \gets$ [$(B(P, \emptyset), \mathbf{Score}(B(P, \emptyset), B_0, W))$]
      \ForAll{queries $q$ \textbf{in} $W$}
        \State $T' \gets$ [ ]
        \ForAll{blueprints $B$ \textbf{in} T}
          \ForAll{engines $e$ \textbf{in} $\{ \mathrm{Aurora}, \mathrm{Athena}, \mathrm{Redshift} \}$}
            \State $B' \gets  (B \cup \{ q \rightarrow e \})$ \Comment{Route query $q$ to $e$ in $B'$}
            \If{$B'$ is valid}
              \State $T' \gets T'$ add $(B', \mathbf{Score}(B', B_0, W))$
            \EndIf
          \EndFor{}
        \EndFor{}
        \LComment{Note that $T'$ is implemented as a top-$k$ heap.}
        \State $T \gets T'$ truncated to the top-$k$ candidates
      \EndFor{}
      \State \Return Best candidate in $T$
    \EndFunction

    \State $B^* \gets$ None
    \ForAll{provisionings $P$ \textbf{near the provisioning in} $B_0$}
      \State $B \gets \Call{DoSearch}{P}$
      \If{$B$ is better than $B^*$}
        \State $B^* \gets B$
      \EndIf
    \EndFor
    \State \Output $B^*$
  \end{algorithmic}
\end{algorithm}
\fi

\ifdefined\BradExtended
\subsection{Triggering Blueprint Optimization}\label{sec:brad-triggers}
\thesystem{} periodically checks a set of \emph{triggers} to decide when to
initiate blueprint optimization.
\thesystem{} initiates blueprint optimization if one of them fires. Concretely,
\thesystem{} uses the following triggers:
\begin{itemize}[leftmargin=*]
  \item \textbf{Aurora / Redshift CPU utilization.} \thesystem{} triggers
  blueprint optimization if they consistently violate preset thresholds
  (e.g., exceeding 85\% or falling below 15\% for 10 minutes or more).
  \item \textbf{Transaction and query latency.} When optimizing for cost under a
  performance SLO, \thesystem{} will trigger re-optimization if these latencies
  consistently exceed the user's SLOs.
  \item \textbf{Recent provisioning change.} If \thesystem{} selects a new
  blueprint with a provisioning change, it will trigger re-optimization after a
  fixed period of time to ensure performance is as expected.
  This is because \thesystem{} makes conservative provisioning decisions to
  avoid selecting blueprints that will violate the user's SLOs. Re-optimizing
  after the new blueprint takes effect gives \thesystem{} an opportunity to
  revisit its choice after observing the workload on the new blueprint.
\end{itemize}
\fi

\subsection{Discussion}
\sparagraph{Blueprint planning practicality.}
Our blueprint planning framework has three practical benefits.
First, blueprints and their scores are \emph{human-interpretable},
making it easy for data engineers to inspect \thesystem{}-chosen designs.
Second, blueprints provide a useful abstraction for realizing cloud
infrastructure designs. Conceptually, they can be ``compiled down'' into
infrastructure-as-code manifests (e.g., CloudFormation~\cite{cloudformation})
for deployment.
Finally, blueprints are generalizable to other cloud infrastructure design
problems that involve cost/performance-based resource provisioning, task
scheduling, and adaptation under changing conditions.
Some example use cases include resource configuration selection for
Ray~\cite{ray18} programs, designing long-lived infrastructures used by Sky
intercloud brokers~\cite{chasins2022sky}, or assembling a model serving
service~\cite{infaas}.

\sparagraph{Adding engines to \thesystem{}.}
\thesystem{} can support additional engines beyond Aurora, Redshift, and Athena.
For an engine to be included in \thesystem{}, it must
\begin{enumerate*}[label=(\roman*)]
  \item support relational data,
  \item have a SQL-based query interface,
  \item expose system metrics (e.g., CPU utilization),
  \item use a deterministic query planner, and
  \item have deterministic operational costs.
\end{enumerate*}
Practically, the engine should also have management APIs that allow \thesystem{}
to programmatically alter its allocated resources (i.e., provisioning) to deploy
blueprints.
By (iv), we mean that the query planner must pick the same physical plan for the
same query if it has the same dataset statistics (e.g., estimated scan
selectivity) (see Section~\ref{sec:brad-bp-scoring-rt}).
Finally by (v), we mean that the cost of running the engine in the cloud must be
a deterministic function of the engine's physical configuration (e.g., its
provisioning and the size of its data) and the user's workload.
For example, the engine's operating cost cannot depend on external factors that
\thesystem{} cannot directly observe (e.g., resource demands from other cloud
users).

\section{Evaluation}\label{sec:evaluation}
In our evaluation, we seek to answer the following questions:
\begin{itemize}[leftmargin=*]
  \item How effective is \thesystem{} at optimizing a data infrastructure for
    cost when compared to serverless autoscaling systems?
    (Section~\ref{sec:eval-e2e-scenarios})
  \item How accurate are the models that \thesystem{} uses to score its
    candidate blueprints and how well do they generalize?
    (Section~\ref{sec:eval-accuracy})
  \ifdefined\BradExtended
    \item How effective is \thesystem{}'s query routing and how much overhead does
      \thesystem{} add to query execution? (Section~\ref{sec:eval-routing})
    \item How effective is \thesystem{}'s blueprint search algorithm?
    (Section~\ref{sec:eval-bp-planner})
    \item How sensitive is \thesystem{}'s blueprint planner to model errors?
      (Section~\ref{sec:eval-sensitivity})
  \fi
\end{itemize}

\noindent
Across five workload scenarios, we find that \thesystem{} selects designs that
meet performance targets while outperforming a serverless Aurora and Redshift
infrastructure and a serverless HTAP system (where comparable) on cost by up to
\BradBestCostSavingsAR{} and \BradBestCostSavingsHtap{} respectively.
\ifdefined\BradVLDB
In our extended paper, we include additional experiments on query routing,
blueprint search, and planning robustness~\cite{brad-extended-tmp}.
\fi

\sparagraph{Overall implementation.}
We implemented \thesystem{} in Python using approximately 30k lines of
code. Although \thesystem{} currently uses AWS services, the concepts underlying
blueprint planning and our scoring models are general and applicable to
other cloud providers.

\subsection{Workload and Experimental Setup}\label{sec:workload}
We evaluate \thesystem{} on a new workload that models the data processing needs
of a fictitious movie theater company called \emph{\company{}}.

\sparagraph{Why create a new workload?}
\thesystem{} automates the design of multi-engine cloud data infrastructures
serving diverse transactional and analytical workloads.
Thus, we need a realistic and diverse workload that warrants multiple
specialized engines. To our knowledge, no such public workloads exist.
The TPC~\cite{tpch,tpcds} and HATtrick benchmarks~\cite{milkai2022good} are
entirely synthetic.
IMDB JOB~\cite{leis2015good} and STATS CEB~\cite{HanWWZYTZCQPQZL21} use
real-world datasets and queries, but only contain OLAP queries as they
are for evaluating query optimizers.
Snowset~\cite{snowflake-nsdi20} has statistics about real OLAP workloads,
but no data or queries.
Our workload addresses these limitations: it
\begin{enumerate*}[label=(\roman*)]
  \item contains transactions and diverse analytical queries,
  \item adapts a real-world dataset, and
  \item mimics Snowset statistics where possible.
\end{enumerate*}

\sparagraph{Dataset.}
We use an adapted version of the IMDB dataset~\cite{leis2015good}, which is
based on real-world data.
As the original IMDB dataset is small (3~GB), we create a larger dataset by
replicating each tuple in the dataset's major tables 30 times. Then, we assign
new values for each replicated primary key and re-assign these values to their
corresponding foreign keys. This approach preserves the dataset's attribute
correlations, skew, and join-key distributions.
We additionally add three synthetic tables representing movie theaters, movie
showings, and ticket orders to capture the company's transactional needs. The
final uncompressed dataset is 160~GB.

\sparagraph{Analytical queries.}
Our workload consists of two classes of analytical queries:
\begin{enumerate*}[label=(\roman*)]
  \item recurring queries (e.g., used for \company{}'s dashboards and
    interactive internal tools), and
  \item complex ad-hoc analytical queries (e.g., representing exploratory data
  analysis).
\end{enumerate*}
The recurring queries consist of single table scans and two table inner joins,
both with predicates.
The complex ad-hoc queries are randomly generated to resemble the IMDB JOB
queries. They span hundreds of distinct templates that join 4 to 15 tables with
complex filter predicates.
Of the unique queries in our workload, 80\% are recurring and 20\% are ad-hoc;
we chose this split to match what our industry partners have observed in their
production workloads.

\sparagraph{Transactions.}
We use 3 transaction types:
\begin{enumerate*}[label=(\roman*)]
  \item \textsc{Purchase},
  \item \textsc{Add\-Showing}, and
  \item \textsc{UpdateMovie}.
\end{enumerate*}
\textsc{Purchase} looks up a theater, selects a showing, inserts a ticket order,
and updates the showing's seat count.
\textsc{AddShowing} looks up a theater and movie and inserts a new showing
record.
\textsc{UpdateMovie} selects a movie from the ``title'' table and edits the
corresponding note column in the ``movie\_info'' and ``aka\_title'' tables.
We run these transactions with a breakdown of 70\% \textsc{Purchase},
20\% \textsc{AddShowing}, and 10\% \textsc{UpdateMovie}.

\sparagraph{Baselines.}
We compare \thesystem{} against
\begin{enumerate*}[label=(\roman*)]
  \item an infrastructure using serverless Aurora for all transactions and
    serverless Redshift for analytics, and
  \item \htap{}, a popular open-source serverless HTAP database.
\end{enumerate*}
Note that these comparisons are not perfectly fair as these systems provide
different guarantees.
We select them because they represent existing industry-standard infrastructure
solutions that provide the same hands-off autoscaling experience.

\sparagraph{Metrics.}
We record three metrics:
\begin{enumerate*}[label=(\roman*)]
  \item transaction latency,
  \item analytical query latency, and
  \item monthly operating cost.
\end{enumerate*}
In our experiments, \thesystem{} optimizes a data infrastructure to minimize
operating cost while ensuring that p90 transaction latency remains under
30~milliseconds and p90 query latency remains under 30 seconds.

\sparagraph{Operating cost calculations.}
We compute \thesystem{}'s operating cost using the on-demand instance hourly
cost scaled to 30 days. For queries running on Athena, we compute their cost
using the reported bytes scanned. We project this value into a monthly cost by
assuming that the query repeats at the same observed rate.
We include storage costs for the tables placed on Aurora and Athena (S3).
For serverless Aurora and Redshift, we use the ACU and RPU values reported by
AWS during the workload and scale them to monthly costs.
For \htap{}, we use the cost reported by the vendor.

\ifdefined\BradExtended
\sparagraph{Considered instance types.}
For Aurora, \thesystem{} currently only considers Graviton-based
instances~\cite{graviton-aurora} and I/O optimized cluster configurations (i.e.,
I/O costs are included in the hourly provisioning cost)~\cite{aurora-ioopt}. We
leave the consideration of different instance hardware types (e.g., r6g vs. r6i
instances) to future work. For Redshift, \thesystem{} considers the dc2 and ra3
family of instance types. 
\fi

\subsection{Optimizing a Data Infrastructure}\label{sec:eval-e2e-scenarios}
We have \thesystem{} optimize a data infrastructure for cost under a
performance constraint in five scenarios faced by \company{}:
\begin{enumerate}[leftmargin=*]
  \item Scaling down an over-provisioned infrastructure.
  \item Scaling engines due to increased load.
  \item Maintaining support for specialized functionality.
  \item Adjusting to user-changed constraints.
  \item Workload intensity variations during a day.
\end{enumerate}

\subsubsection{Scaling Down an Over-Provisioned Infrastructure}
\label{sec:eval-scale-down}
\begin{figure}
  \begin{overpic}[width=0.95\columnwidth]{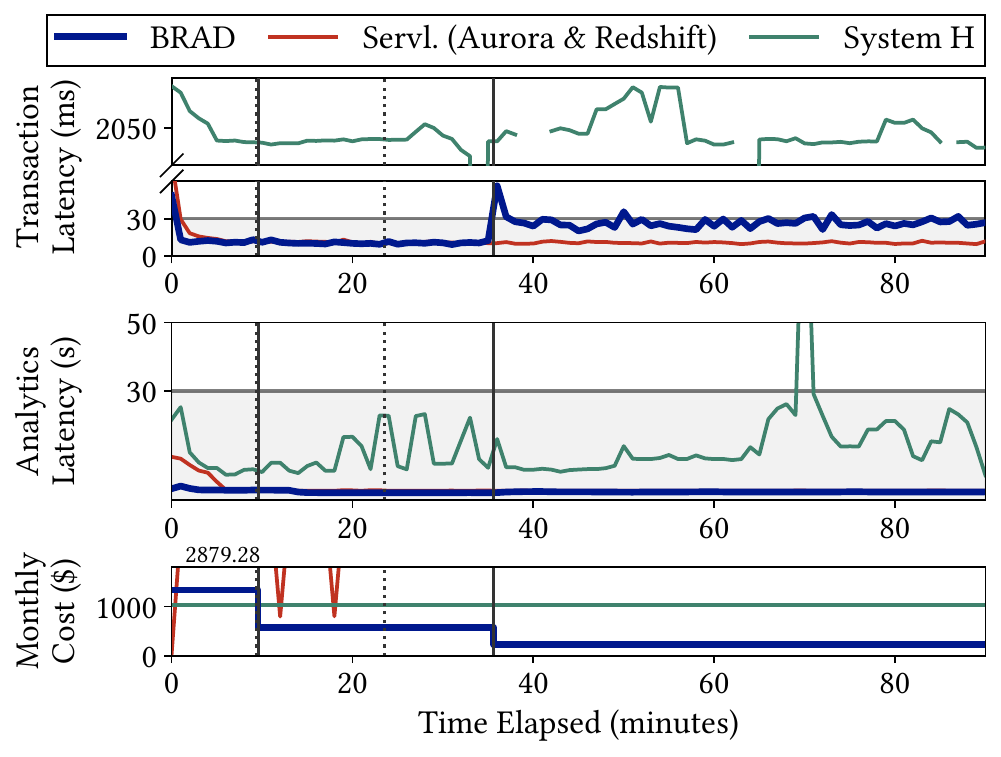}
    \put(21,53.5){\captionlabel{A}}
    \put(41,15.5){\captionlabel{B}}
    \put(44.5,53.5){\captionlabel{C}}
    \put(65,13){\captionlabel{D}}
    \put(77,29){\captionlabel{E}}
    \put(86,54.5){\captionlabel{F}}
  \end{overpic}
  \vspace{-0.75em}
  \caption{\thesystem{} reduces cost while maintaining p90 latency constraints
  (shaded region). The dotted (solid) vertical lines indicate when a new
  blueprint is chosen (takes effect).}
  \label{fig:e2e-scale-down}
  \vspace{-0.5em}
\end{figure}
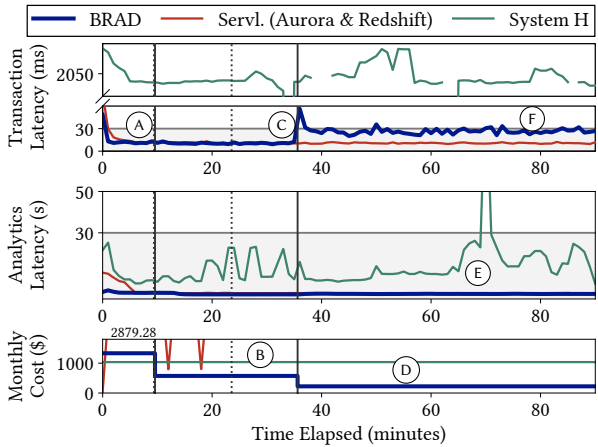

\company{} has been struggling with their data infrastructure. After learning
about \thesystem{}, they adopt it to free up their data engineers.
\company{} deploys \thesystem{} on their infrastructure, consisting of two
Aurora db.r6g.xlarge instances (primary and read replica) and two dc2.large
Redshift nodes. Following conventional wisdom, they use Redshift for analytical
queries and Aurora for transactions.
Figure~\ref{fig:e2e-scale-down} shows workload performance and the monthly
operating cost over time. The shaded area indicates \company{}'s performance
constraints: 30~ms p90 transaction latency and 30~s p90 analytics latency.

Soon after starting, \thesystem{} triggers blueprint optimization \encircle{A}
because it detects a low Redshift CPU utilization.
\thesystem{} removes the Aurora read replica, shifts the entire analytical
workload onto Aurora, and pauses Redshift. \thesystem{} makes these changes to
reduce cost \encircle{B}, as it correctly predicts that Aurora is sufficient to
handle the workload.
After observing the workload on this new blueprint, \thesystem{} then correctly
predicts that Aurora can support the workload with a smaller (cheaper) instance
type and thus downscales Aurora to two db.t4g.medium instances to further reduce
cost \encircle{D}.
The momentary spike in p90 transactional latency is due to the Aurora primary
failover that occurs when changing instance types \encircle{C}.
The chosen blueprint meets \company{}'s performance constraints \encircle{E}
\encircle{F}.
From these results, we draw the following two conclusions.

\textbf{\thesystem{} reduces operating cost by \ScaleDownBradSelfSavings{} over
its starting provisioning and by \ScaleDownBradHtapSavings{} over \htap{}, the
next best baseline.}
Serverless Aurora and Redshift is \ScaleDownBradARSavings{} more expensive than
\thesystem{}'s because serverless Redshift has a large minimum size, making it
cost-ineffective on this workload. Although \htap{} is only
\ScaleDownBradHtapSavings{} more expensive than \thesystem{}, its p90
transaction latency is nearly 100$\times$ higher than the other baseline. We
hypothesize that this is due to \htap{}'s internal replication on writes.

\textbf{\thesystem{} shifts workloads across engines to reduce cost,
differentiating it from static multi-engine autoscaling infrastructures.}
\thesystem{} correctly predicts that Aurora can support the analytical workload,
enabling it to pause Redshift to reduce cost. This decision would never be
considered in static autoscaling infrastructures, such as our serverless Aurora
and Redshift baseline, since they only scale to respond to system load while
keeping the workload assignment fixed (i.e., analytics always run on Redshift).

\subsubsection{Scaling Engines Due to Increased Load}
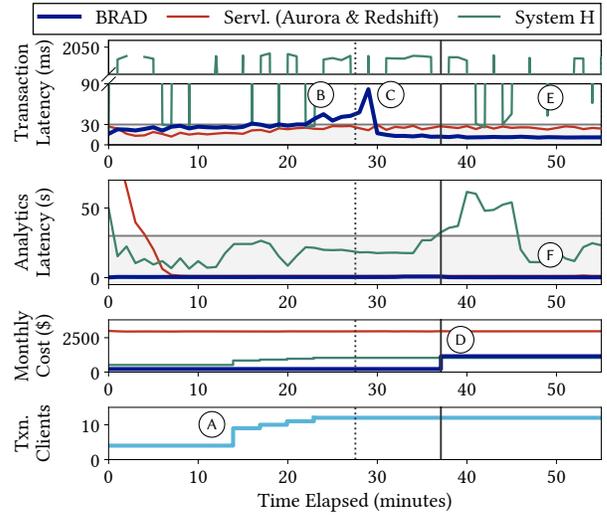
\begin{figure}[t]
  \begin{overpic}[width=0.95\columnwidth]{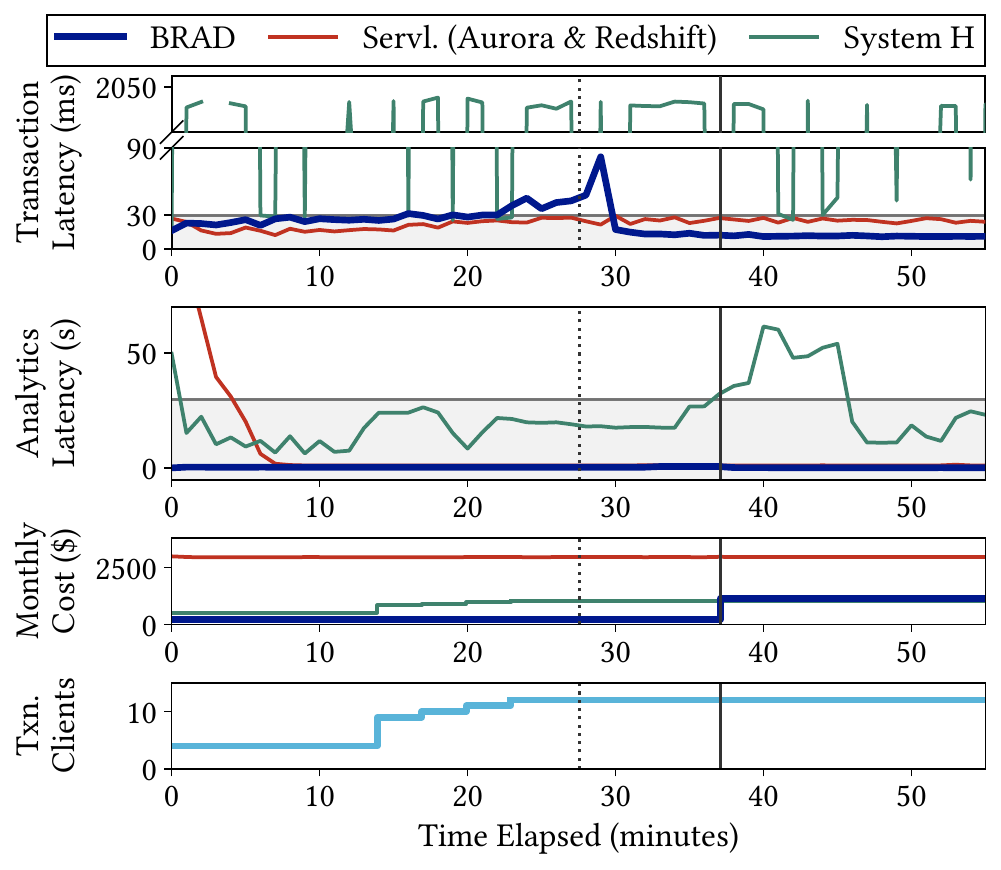}
    \put(32,15){\captionlabel{A}}
    \put(50,69.5){\captionlabel{B}}
    \put(61.5,69.5){\captionlabel{C}}
    \put(73,29){\captionlabel{D}}
    \put(88,69){\captionlabel{E}}
    \put(88,43){\captionlabel{F}}
  \end{overpic}
  \vspace{-0.75em}
  \caption{As transactional load increases, \thesystem{} switches to
    a larger Aurora instance and also removes the read replica.}
  \label{fig:e2e-scale-up-txn}
  \vspace{-0.5em}
\end{figure}

As \company{} grows, their transactional load increases.
Figure~\ref{fig:e2e-scale-up-txn} shows how \thesystem{} handles this change; we
plot the same metrics as before and include the number of
transactional clients over time \encircle{A}. We begin with the same
blueprint as the end of the previous scenario.
After a few minutes, \thesystem{} notices that the transaction p90 latency is
exceeding \company{}'s 30~ms ceiling \encircle{B} and triggers blueprint
optimization.
\thesystem{} chooses to scale up Aurora to a single db.r6g.xlarge instance as it
correctly predicts that a single Aurora instance can support both the increased
transactional load and the running analytical queries.
The spike in p90 transactional latency \encircle{C} is when the Aurora primary
failover occurs. On the new blueprint, the transactional p90 latency falls under
the latency ceiling \encircle{E}; the analytical queries also continue to
complete under the 30~s ceiling \encircle{F}.
Again, the increased \htap{} analytics latency might be due to a combination of
its internal physical autoscaling and storage writes.

\textbf{This shows that \thesystem{} reacts to transactional
load to maintain latency constraints.} The blueprint that \thesystem{} selects
is \ScaleUpTxnBradARSavings{} cheaper than the serverless Aurora and Redshift
baseline \encircle{D} but up to \ScaleUpTxnBradHtapDiff{} more expensive than \htap{}.
Serverless Aurora and Redshift is more expensive due to Redshift's large minimum
size.

\subsubsection{Maintaining Support for Specialized Functionality}\label{sec:eval-specialized}
To increase engagement, \company{} decides to deploy a new feature that
recommends movies similar to the ones shown in their theaters.
To make recommendations, they use \emph{vector similarity
search}~\cite{rag2020,pgvector} queries that find movies with title embeddings
that are closest to a given movie's title embedding.
\company{} deploys this feature on their existing infrastructure; since
similarity search is only supported on Aurora, they place the relevant tables on
Aurora and run all other queries that access these tables on Aurora as well.
They use two Aurora db.r6g.2xlarge instances and two dc2.large Redshift nodes.

\begin{figure}[t]
  \begin{overpic}[width=0.95\columnwidth]{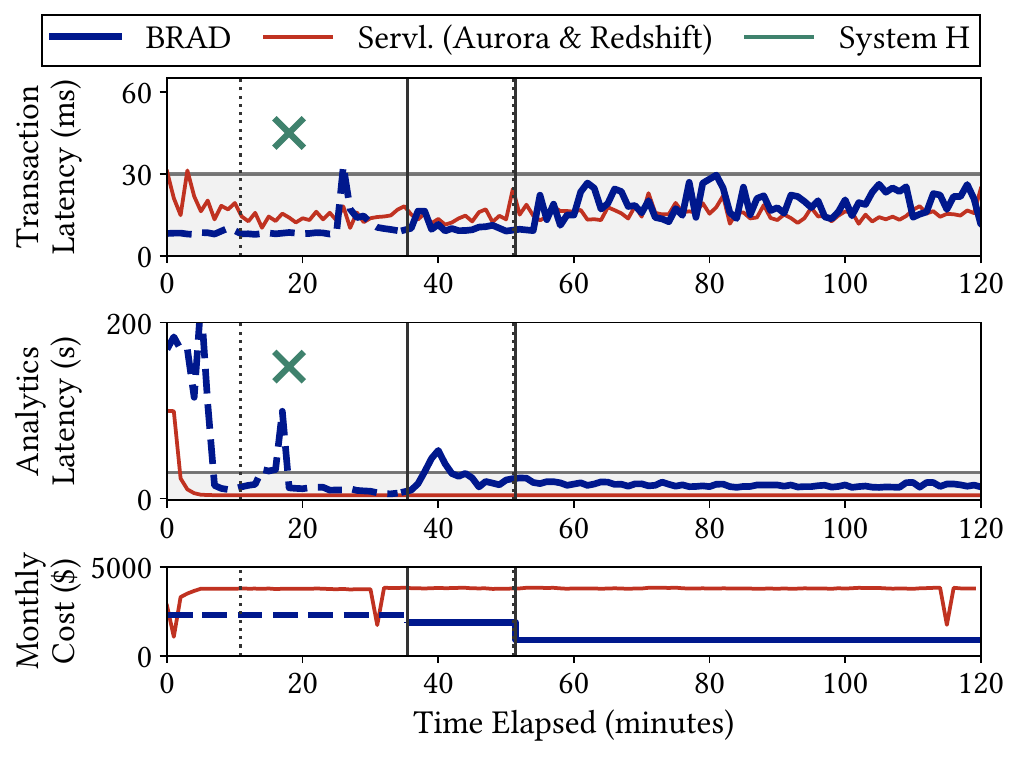}
    \put(22,61){\captionlabel{A}}
    \put(17,30){\captionlabel{B}}
    \put(42,33){\captionlabel{C}}
    \put(80,30){\captionlabel{D}}
    \put(88,59){\captionlabel{E}}
    \put(86,13){\captionlabel{F}}
  \end{overpic}
  \vspace{-0.75em}
  \caption{\thesystem{} runs vector similarity search on Aurora and shifts other
  queries to Redshift for performance. \htap{} does not support vector
  similarity search.}
  \label{fig:e2e-specialized}
\end{figure}

Figure~\ref{fig:e2e-specialized} shows performance and
infrastructure cost over time. The dashed lines are from before \thesystem{}
deploys its first optimized blueprint. Crucially, \htap{} \emph{cannot} run this
workload because it does not support vector similarity search \encircle{A}.
\thesystem{} notices that the analytical p90 latency exceeds \company{}'s
constraint of 30~s \encircle{B} and initiates blueprint optimization.
\thesystem{} selects a blueprint that shifts the non-vector similarity search
queries onto Redshift (replicating the tables they access onto Redshift) while
keeping the vector similarity search queries on Aurora.
It also correctly predicts that it can downscale Aurora (to a db.r6g.xlarge
instance) to save cost, as much of Aurora's former query load was moved onto
Redshift.
After making this change, the workload's performance falls within the user's
performance constraints \encircle{D} \encircle{E}.
The momentary spike in analytics latency at the 40 minute mark \encircle{C} is
due to a cold Redshift cache when \thesystem{} first moves queries onto Redshift.
The serverless Aurora and Redshift design is \SpecializedBradARSavings{} more
expensive \encircle{F} because of the large Redshift minimum size
and because Aurora has scaled up to support the new similarity search queries.

\textbf{This scenario shows how \thesystem{} is fundamentally different from
single-system (e.g., HTAP) solutions like \htap{}.}
When using a single system to run a diverse data workload, you can always run
into situations where you want to use a feature that does not exist on your
system of choice.
In contrast, in the \thesystem{} architecture, one can (in concept) always
\emph{incorporate} a system with the required functionality into the underlying
infrastructure.

\subsubsection{Adjusting to Changed Constraints}
\begin{figure}
  \begin{overpic}[width=0.95\columnwidth]{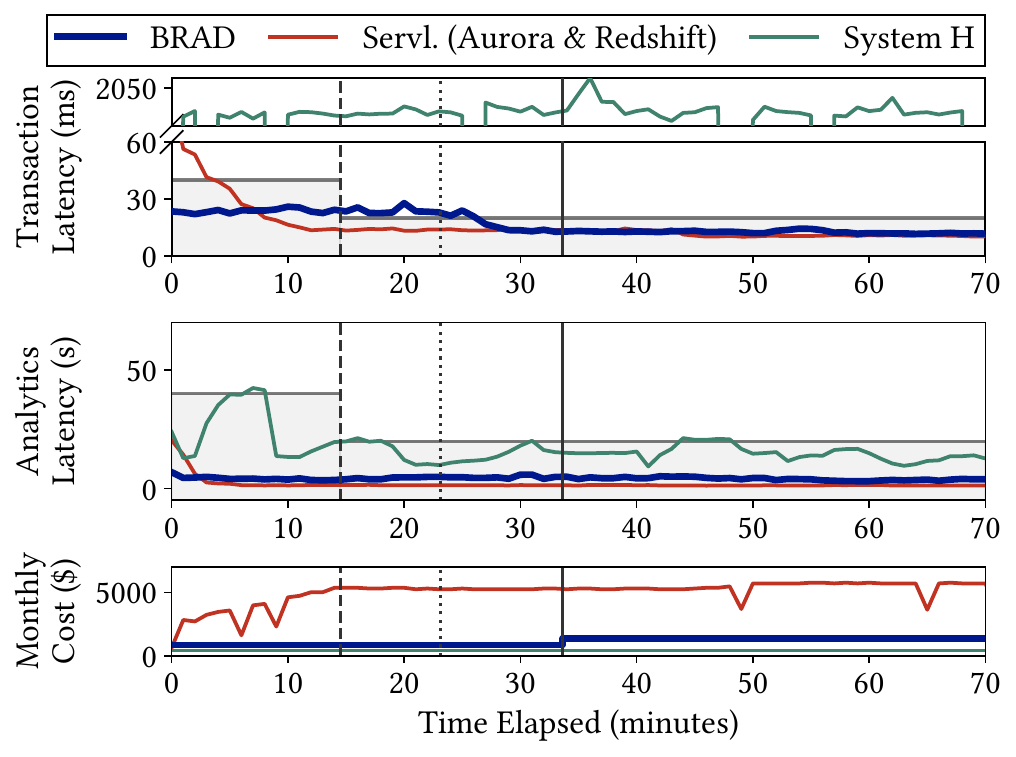}
    \put(18,38){\captionlabel{A}}
    \put(24,56.5){\captionlabel{B}}
    \put(29,38){\captionlabel{C}}
    \put(39,55){\captionlabel{D}}
    \put(56,14){\captionlabel{E}}
    \put(58,54){\captionlabel{F}}
    \put(63,33){\captionlabel{G}}
  \end{overpic}
  \vspace{-0.75em}
  \caption{After the user changes their SLO constraints, \thesystem{} selects a
  new blueprint to meet the new constraints.}
  \label{fig:slo}
  \vspace{-0.5em}
\end{figure}
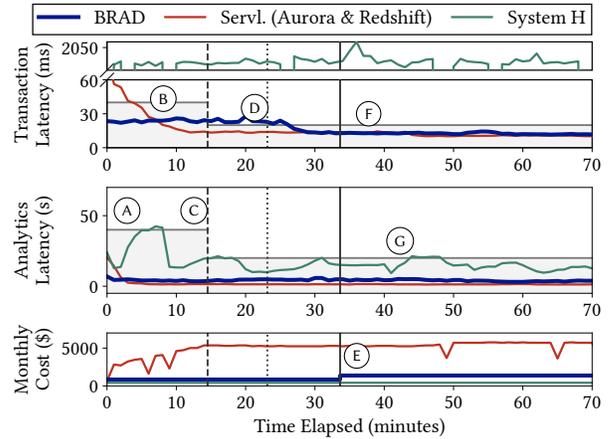

As \company{}'s business changes, they revise their performance constraints;
Figure~\ref{fig:slo} shows how \thesystem{} adapts to their new needs.
\company{} initially uses a p90 transaction and query SLO of 40~ms and 40~s
respectively \encircle{A} \encircle{B}.
\thesystem{} starts with one db.r6g.xlarge Aurora instance and two Redshift
dc2.large nodes.
Later, \company{} lowers their transaction and query SLOs to 20~ms and 20~s
respectively (the change happens at the dashed line in Figure~\ref{fig:slo}
\encircle{C}).
This SLO change causes the transaction latency to exceed \company{}'s
constraints.
Thus, \thesystem{} scales up Aurora to one db.r6g.2xlarge instance \encircle{D} and
leaves Redshift as-is. This change increases the operating cost \encircle{E} as
\thesystem{} switches to a larger Aurora instance.
After \thesystem{} finishes transitioning the infrastructure, the transaction
latency falls under the new 20~ms p90 constraint \encircle{F}. The query latency
constraint also remains under the new 20~s p90 constraint \encircle{G}.
\thesystem{}'s operating costs are \SLOBradSavingsOverAR{} lower than the
serverless Aurora and Redshift baseline. This is because serverless Redshift has
a large minimum size.
While \htap{} ends with a \SLOHtapSavingsOverBrad{} lower monthly operating cost
than \thesystem{}, it does not meet the 20~ms transaction latency constraint
(its transactions have a latency around 2 seconds).
We again think that this elevated latency is due to \htap{}'s internal
replication on writes.
This scenario shows that \thesystem{} adapts to changes to a user's constraints.

\subsubsection{Workload Intensity Variations During a Day.}\label{sec:eval-e2e-day}
\begin{figure}[t]
  \begin{overpic}[width=0.95\columnwidth]{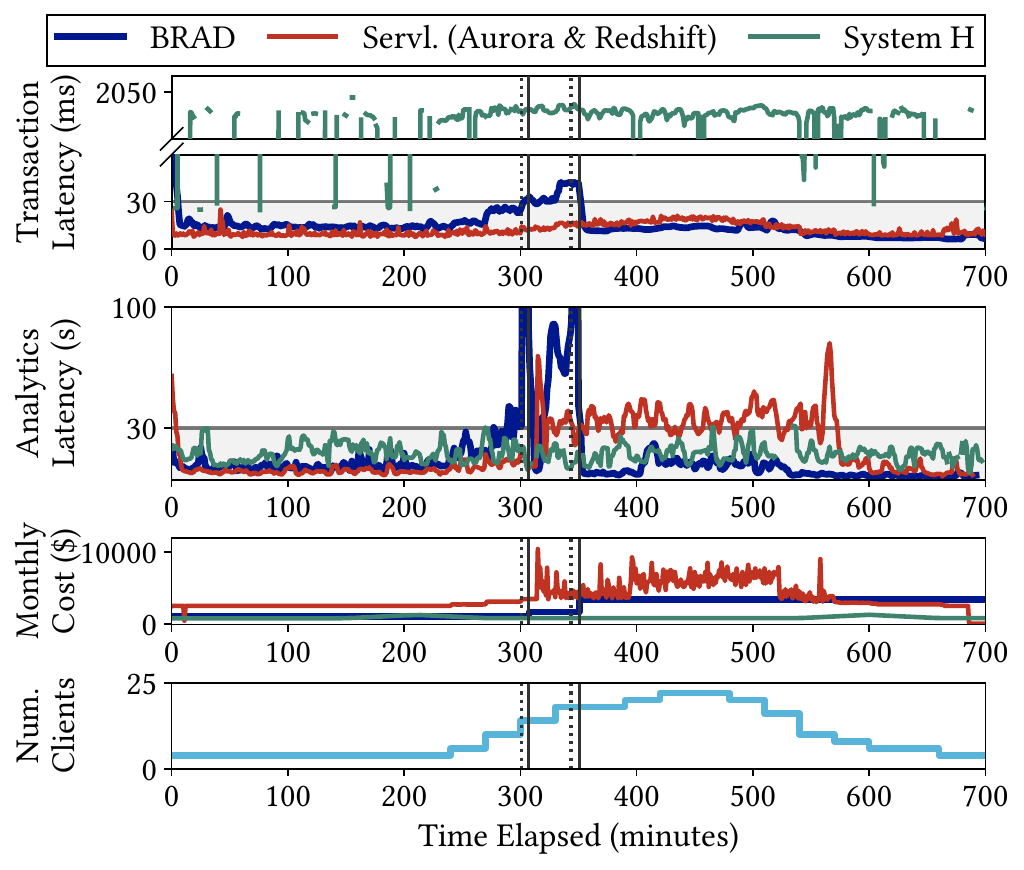}
    \put(26,28){\captionlabel{A}}
    \put(45,46){\captionlabel{B}}
    \put(52,68){\captionlabel{C}}
    \put(65,48){\captionlabel{D}}
    \put(67,32){\captionlabel{E}}
    \put(65,13){\captionlabel{F}}
  \end{overpic}
  \vspace{-0.75em}
  \caption{\thesystem{} optimizes for load variations during the day.}
  \label{fig:e2e-daylong}
  \vspace{-0.5em}
\end{figure}
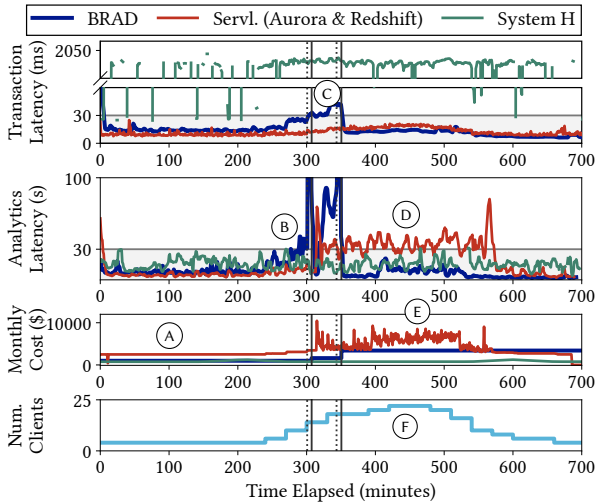

Finally, we run \thesystem{} on a workload representing a full day at
\company{}.  For practical reasons, we scale the actual workload to 12 hours.
Figure~\ref{fig:e2e-daylong} shows performance and cost over the day.
We use the workload and dataset from Section~\ref{sec:workload} adapted to mimic
the Snowset trace~\cite{snowflake-nsdi20}. Concretely, we run queries with a
run time distribution similar to the Snowset trace and vary the number of
clients issuing queries and transactions to mimic the diurnal pattern observed
in Snowset (a peak near the middle of the day, Figure~\ref{fig:e2e-daylong}
\encircle{F}).

Initially, the workload is light. \thesystem{} uses a blueprint with four
dc2.large Redshift nodes and two Aurora db.t4g.medium instances, which is
\BRADDayLongInitialCostImprovement{} cheaper than the serverless Aurora and
Redshift baseline \encircle{A}.
As the workload intensity increases, \thesystem{} detects the increases in
latency \encircle{B} \encircle{C} and triggers blueprint optimization,
ultimately scaling Redshift up to 16 nodes and Aurora to one db.r6g.xlarge
instance at the peak.
The serverless Aurora and Redshift baseline also scales up, but it does not
consistently meet the analytics performance target of 30~s \encircle{D}, despite
being \BRADDayLongPeakCostImprovement{} more expensive than \thesystem{}'s
design \encircle{E} at the workload peak.
\htap{} maintains the p90 analytics latency SLO throughout the workload, but its
transactional latency is again almost two orders of magnitude higher than the
other systems (we hypothesize for the same reasons as discussed earlier).
The brief analytics latency spikes are due to Redshift resizes, which force
clients to reconnect.

\textbf{This result shows that \thesystem{} effectively responds to
load variations during the day}.
Over the day, \thesystem{} maintains performance targets while reducing
cumulative cost by \BRADDayLongCumulativeCostImprovement{} compared to the
serverless Aurora and Redshift baseline.

\subsection{Scoring: Model Accuracy and Generalization}\label{sec:eval-accuracy}
We next examine the test accuracy of our predictive models and their
generalizability across common workload shifts.

\subsubsection{Model Accuracy}
Table~\ref{tbl:overall-test-error} shows the median test Q error of our models
for each engine. Q error is $Q(p, a) = \mathrm{max}(p/a, a/p)$, where $p$ refers
to the predicted value and $a$ to the actual value. Lower is better; 1 is the
best possible score.
We train each run time and data accessed model using approximately 8000 queries,
validate on 2000 queries, and test on 125 unseen queries. Our query dataset
consists of over 1000 unique join templates.
The models for Athena perform better than Aurora and Redshift because the run
time distribution of Athena queries has a lower variance.
We test our provisioning and transaction latency models on an unseen
provisioning that is larger (i.e., has more resources) than all the training
provisionings.
Overall, we find that our models' prediction accuracy is sufficient for
\thesystem{} to design effective infrastructures
(Section~\ref{sec:eval-e2e-scenarios}).

\begin{table}[t]
    \footnotesize
    \centering
    \caption{Median test Q error of our blueprint scoring models.}
    \vspace{-0.75em}
    \begin{tabularx}{\columnwidth}{Xlll}
      \toprule
      \textbf{Prediction Target} & \textbf{Aurora} & \textbf{Redshift} & \textbf{Athena} \\
      \midrule
      Query Run Time & 1.5769 & 1.6539 & 1.3427 \\
      Data Accessed & {\color{lightgray} --} & {\color{lightgray} --} & 1.2614 \\
      \midrule
      Run Time on Different Provisioning & 1.6718 & 1.6824 & {\color{lightgray} --} \\
      Transaction Latency & 1.2030 & {\color{lightgray} --} & {\color{lightgray} --} \\
      \bottomrule
    \end{tabularx}
    \label{tbl:overall-test-error}
\end{table}

\subsubsection{Generalizability}
We evaluate our query run time model's generalizability on three workload
shifts:
\begin{enumerate*}[label=(\roman*)]
  \item unseen join templates,
  \item adding a new table to the dataset, and
  \item a larger dataset size.
\end{enumerate*}

\begin{figure}
  \centering
  \includegraphics[width=0.98\columnwidth]{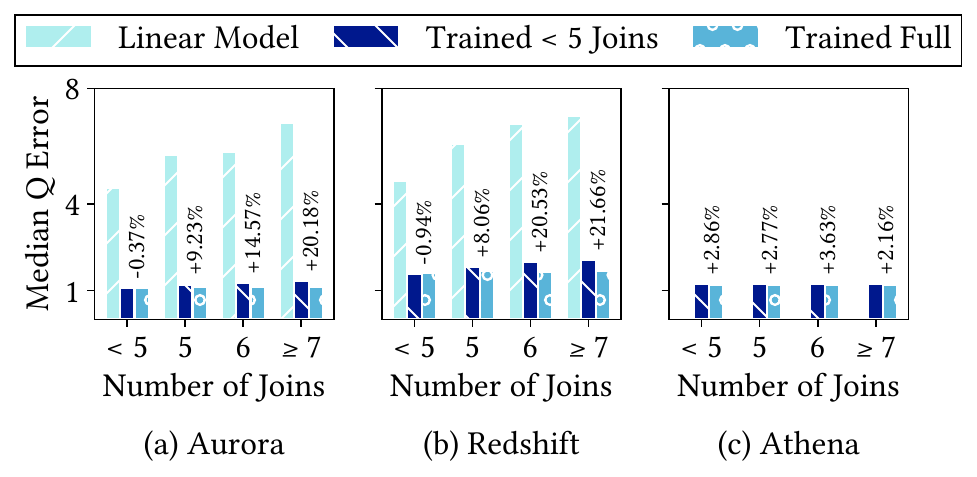}
  \vspace{-0.75em}
  \caption{\thesystem{} generalizes to unseen join templates.}
  \label{fig:eval-rt-gen-template}
  \vspace{-0.5em}
\end{figure}

\sparagraph{Unseen join templates.}
We train our run time models on queries with less than 5 joins. We then test the
model's predictions on queries with 5, 6, and $\geq$ 7 joins.
Figure~\ref{fig:eval-rt-gen-template} shows each model's median test Q error
compared with
\begin{enumerate*}[label=(\roman*)]
  \item a model trained on all the join templates (``full''), and
  \item a na\"ive linear model that scales the cost returned by the engine's query
  optimizer to a run time.
\end{enumerate*}
We label the percentage difference from the model trained on the full dataset.
Our model generalizes across unseen join templates with a Q error of at most
20\% above the model trained on the full dataset.
Our model still performs much better than the na\"ive linear model, which has a
Q error of at least 4.6. Since Athena's optimizer does not provide a query cost,
we do not include a linear model result.

\begin{figure}
    \centering
    \includegraphics[width=0.98\columnwidth]{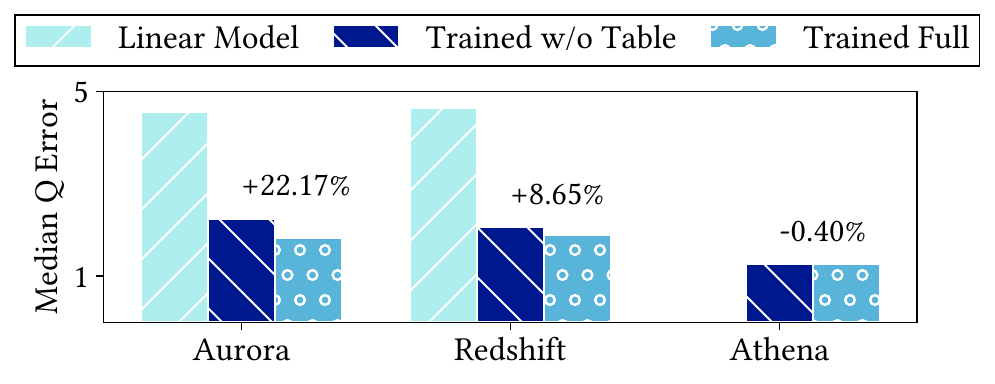}
    \vspace{-1em}
    \caption{\thesystem{} generalizes to an added table.}
    \label{fig:eval-rt-gen-table}
    \vspace{-0.5em}
\end{figure}

\sparagraph{Added table.}
We train our run time models on queries that do not access the ``person\_info''
table and test them only on queries that access the ``person\_info'' table.
This table is around 10~GB (the second largest table in the dataset); the
overall dataset size is 160~GB.
Figure~\ref{fig:eval-rt-gen-table} shows our results using the same baselines
and notation as Figure~\ref{fig:eval-rt-gen-template}.
Our model generalizes to an added table with a Q error at most 22\% above the
model trained on the full dataset.
The linear model still does poorly, with a Q error of 4.6.

\begin{figure}
    \centering
    \includegraphics[width=0.98\columnwidth]{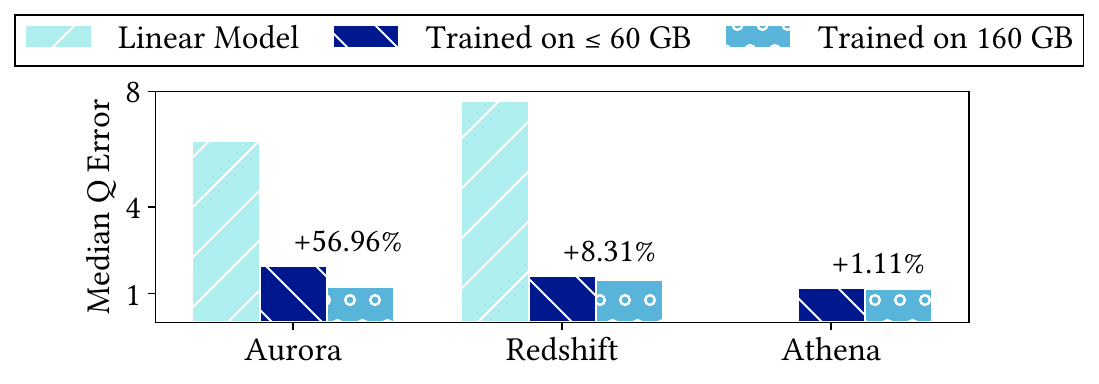}
    \vspace{-0.75em}
    \caption{\thesystem{} generalizes to an increased dataset size.}
    \label{fig:eval-rt-gen-size}
    \vspace{-0.5em}
\end{figure}

\sparagraph{Increased dataset size.}
Finally, we evaluate our run time model's generalizability to larger datasets.
We train our models using queries executed on a 3 GB, 20 GB, 40 GB, and 60 GB
version of our workload dataset. Then we test the models on the original 160~GB
dataset.
Figure~\ref{fig:eval-rt-gen-size} shows that our model generalizes to the larger
dataset with a Q error of 1.1\%, 8.3\%, and 57\% above a model specifically
trained on the 160 GB dataset on Athena, Redshift, and Aurora respectively.
The Aurora model has a higher error due to a change in caching behavior that
occurs beyond 60 GB.
The linear model has a Q error of 6.3 and 7.7 on Aurora and Redshift
respectively. Overall, these results are sufficient for \thesystem{} since an
increase from 60~GB to 160~GB would likely happen over a longer period of
time, allowing for \thesystem{} to update its models given newly observed data.

\ifdefined\BradExtended
\subsection{Query Routing Quality and Overhead}\label{sec:eval-routing}
Next, we evaluate \thesystem{}'s query routing
(Section~\ref{sec:routing-key-idea}) against four baselines:
\begin{enumerate*}[label=(\roman*)]
  \item selecting an engine randomly,
  \item routing all queries to Redshift,
  \item routing using the \thesystem{} run time model (Section~\ref{sec:brad-bp-scoring-rt})
    but excluding its inference overhead, and
  \item routing using the \thesystem{} run time model.
\end{enumerate*}
We use 125 queries from our workload;  80\% of the queries represent
recurring queries and 20\% are complex ad-hoc queries.
We report the geomean slowdown over optimal per routing
decision (i.e., lower is better, 1.0$\times$ is optimal). That is, for each
query, we divide its run time on the chosen engine over its run time
on the fastest engine, and take the geomean across queries.

Figure~\ref{fig:routing-quality} shows our results.
\thesystem{} performs the best with a geomean slowdown of
\BRADRouting{}, comparable to using the run time model and excluding
its inference overhead (\RTOnlyRouting{}). Routing by actually using the run
time model (i.e. including its inference overhead) has a geomean slowdown of
\RTWithOverheadRouting{}.
These results highlight why we do not directly use our run time model for
routing; it imposes a high overhead on the query's critical path (up to 115~ms),
 negatively affecting the routing performance. They also show
that routing needs an intelligent strategy; the Random (\RandomRouting{}) and
Redshift Only (\RedshiftOnlyRouting{}) strategies both perform worse than
\thesystem{}.

\begin{figure}
    \centering
    \includegraphics[width=\columnwidth]{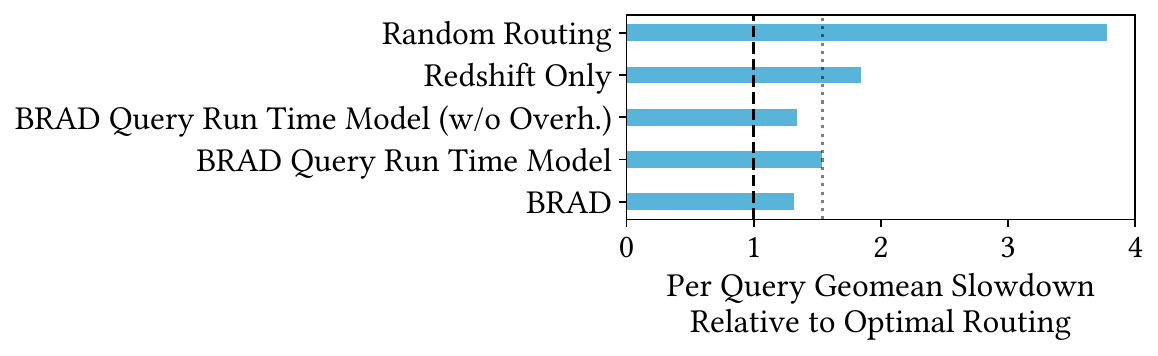}
    \vspace{-2em}
    \caption{\thesystem{}'s routing quality is on par with its query run time
      model but without the inference overhead.}
    \label{fig:routing-quality}
    \vspace{-1em}
\end{figure}

\sparagraph{Overall query processing overhead.}
Similar to common database
proxies~\cite{tigger-butrovich23,rds-proxy,pgbouncer}, \thesystem{} imposes some
overhead. We measure a median overhead of \BRADTxnOverhead{} per complete
transaction (these are interactive multi-statement transactions) and around
\BRADAnaOverhead{} per analytical query. These could be further reduced, as
\thesystem{} is now implemented in Python, but we believe they are reasonable
given that \thesystem{} operates in the cloud, serving remote clients. 
\fi

\ifdefined\BradExtended
\subsection{Blueprint Search Effectiveness}\label{sec:eval-bp-planner}
We next evaluate the effectiveness of \thesystem{}'s blueprint search algorithm.
To compare the search algorithms, we report the final scalar score computed by
\thesystem{}'s optimizer (Section~\ref{sec:brad-bp-comparator}). All baselines
use the same scoring models and optimize for the same workload; they only differ
in how they search for candidate blueprints.

We compare against three baselines:
\begin{enumerate*}[label=(\roman*)]
  \item uniform random sampling,
  \item na\"ive greedy, and
  \item exhaustive search.
\end{enumerate*}
In uniform random sampling, each query is mapped to an engine that is chosen
uniformly at random. We repeat this process to sample 10,000 blueprints and
select the best one.
In naïve greedy, each query is mapped to the engine with the lowest predicted
run time (Section~\ref{sec:brad-bp-scoring-rt}) without consideration of any
other queries.
Finally, exhaustive search looks through all possible mappings and therefore has
a run time that is exponential in the number of queries. To be tractable, we use
only 12 randomly chosen queries from the IMDB workload, comprising a search
space of 530,000 candidates.

We compare the algorithms on the scale down scenario from
Section~\ref{sec:eval-scale-down}. Figure~\ref{fig:planner-baselines} shows that
\thesystem{}'s beam search finds a blueprint as good as exhaustive search. The
score represents a monetary cost, and so lower is better. \thesystem{}'s
blueprint's score is significantly lower than the blueprints selected through
uniform sampling and the naïve greedy approach. This result indicates that
\thesystem{}'s beam strategy is effective at finding optimized blueprints.

\begin{figure}[t]
  \centering
  \includegraphics[width=0.8\columnwidth]{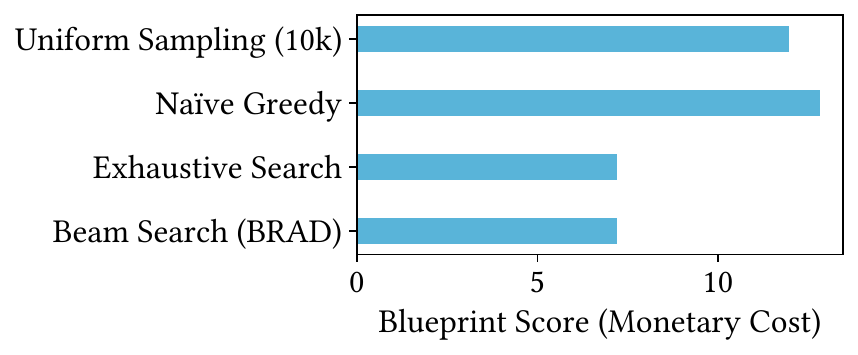}
  \vspace{-1em}
  \caption{\thesystem{}'s blueprint planner compared to baselines.}
  \label{fig:planner-baselines}
\end{figure}
\fi

\ifdefined\BradExtended
\begin{figure}[t]
    \centering
    \includegraphics[width=\columnwidth]{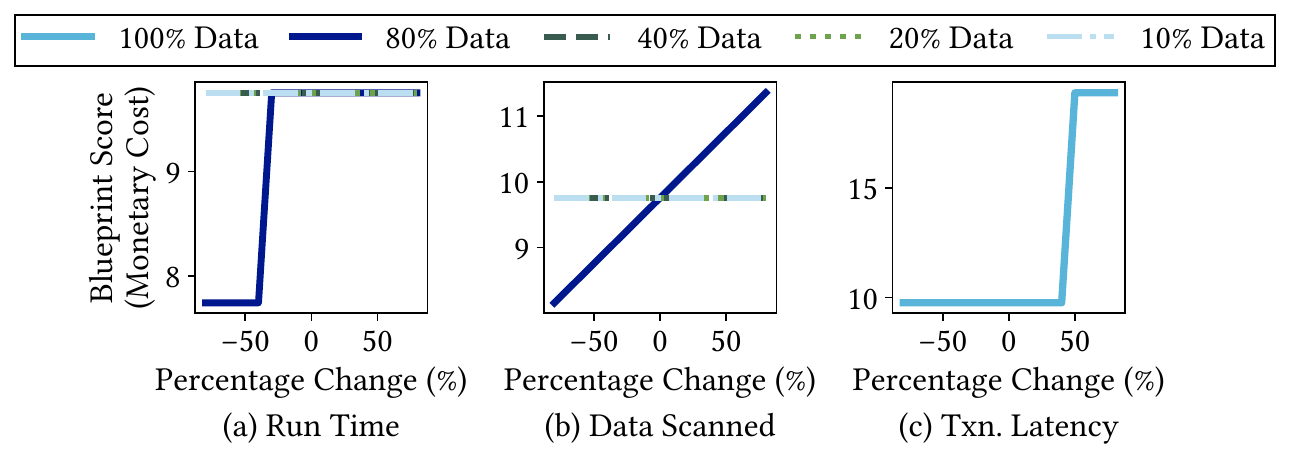}
    \vspace{-1.5em}
    \caption{\thesystem{}'s planner is robust to prediction errors.}
    \label{fig:planner-sensitivity}
    \vspace{-0.5em}
\end{figure}

\subsection{Blueprint Planner Sensitivity}\label{sec:eval-sensitivity}
Finally, we examine our blueprint planner's sensitivity to prediction errors.
We inject errors into \thesystem{}'s predictions during blueprint planning and
record the selected blueprint's score. Concretely, we select a random subset of
the predicted values (query run time, data scanned, and transaction latency) and
increase (or decrease) their predicted values by a percentage. We use subsets
that include 10\%, 20\%, 40\%, and 80\% of the predictions. We run the blueprint
planner on the scale down scenario from Section~\ref{sec:eval-scale-down}.
Figure~\ref{fig:planner-sensitivity} shows our results. The $x$-axis is the
amount of injected error, which we vary from $-80\%$ to $+80\%$. The $y$-axis is
the blueprint's scalar score.
We study the effects of prediction error on query run time, data scanned, and
transaction latency.

Figure~\ref{fig:planner-sensitivity}(a) shows our results for query run time.
There is no change in the selected blueprint even when up to 40\% of the
predicted query run times have injected errors of $\pm 80\%$. When 80\% of the
predictions have injected errors of more than $-40\%$, \thesystem{} selects a
different (cheaper) blueprint as it predicts that the cheaper blueprint can meet
the performance constraints.

Figure~\ref{fig:planner-sensitivity}(b) shows results for a query's predicted
data scanned. Similar to run times, there is no change to the chosen blueprint
when up to 40\% of the predictions have injected errors. At 80\%, \thesystem{}
chooses to route queries onto Athena. Thus the blueprint's monetary cost varies
linearly with respect to the injected error.

Figure~\ref{fig:planner-sensitivity}(c) shows results for transaction latency.
Note that we inject errors into 100\% of the data since \thesystem{} makes just
one latency prediction. Here, an injected error of $+50\%$ causes \thesystem{}
to select a larger Aurora instance (to meet the latency constraint), which
increases the blueprint score (monetary cost).

Overall, our results indicate that \thesystem{}'s planner is robust to
prediction errors; more than 40\% of the predictions need to have an error of
more than $\pm 50\%$ for \thesystem{} to choose a different blueprint. The
intuition is that blueprints represent coarse-grained design decisions, and thus
are more tolerant to prediction errors.
\fi

\section{Related Work}\label{sec:related-work}

\sparagraph{Instance-optimized, self-driving, and auto-tuning systems.}
Recent work has proposed techniques to automatically
\begin{enumerate*}[label=(\roman*)]
  \item adapt data systems to the workload~\cite{sagedb-cidr-19,marcus2022bao,marcus2019neo,
  krishnan2018learning,yu2020reinforcement,marcus2018deep,tsunami,flood,jialin2021mto,
  learnedindexes,yu2022treeline,perron2023cackle,vetl, proteus-abebe22},
  \item manage complex systems~\cite{lim22, pavlo17, pavlo19, pavlo21, keebopaper, AutoWLM}
  \item adapt cloud database instance sizing~\cite{venkataraman2016ernest,
  ortiz2018slaorchestrator, ortiz2016perfenforce, ortiz2019performance}, and
  \item tune their knobs~\cite{tuning2017dana,ottertune,llamatune}.
\end{enumerate*}
In contrast, \thesystem{} optimizes an entire multi-engine data infrastructure
instead of tuning individual services.
\thesystem{} can be seen as applying instance-optimization at the
granularity of cloud database services instead of within a database
engine~\cite{sagedb-cidr-19}.

\sparagraph{Simplifying and optimizing the cloud.}
Like \thesystem{}, recent research has explored ways to simplify and optimize
the design and operation of cloud infrastructures. These thrusts include
\begin{enumerate*}[label=(\roman*)]
  \item high-level cloud programming
    abstractions~\cite{cheung2021new,hydroflow,ssms,ray18},
  \item infrastructure as code~\cite{cloudformation,terraform},
  \item enhancing cross-cloud compatibility~\cite{chasins2022sky,sky-website}, and
  \item improving resilience across services~\cite{li2023darq}.
\end{enumerate*}
\thesystem{}'s key difference is that it focuses on simplifying cloud
infrastructures containing multiple relational database systems while optimizing
their use for cost under a performance constraint.

\sparagraph{Single-system solutions.}
Another way to handle diverse data workloads is to use a single specialized
(e.g., HTAP) database system designed for high performance across many
workloads~\cite{HANAoverview, TiDBpaper, HyPer11, oracle15, Hana12}.
For some workloads (e.g., real-time analytics), such systems can be more
efficient than \thesystem{} because they are not internally constrained by
engine boundaries.
But these single-system solutions can be difficult to migrate to and they limit
users to their specific feature set.
In contrast, \thesystem{} is designed to optimize existing multi-engine data
infrastructures and (in concept) can include new systems to support specialized
functionality (Section~\ref{sec:eval-specialized}).

\sparagraph{Polystores and federated databases.}
Prior work on polystores~\cite{duggan15bigdawg, wang2017myria,
alotaibi2019towards, vogt2018polypheny, zheng2021awesome, agrawal2018rheem,
podkorytov2019hybridpoly} and federated databases~\cite{breitbart1992overview,
breitbart1988update, hwang1994myriad, pu1988superdatabases, sheth1990federated,
georgakopoulos1991multidatabase, josifovski2002garlic, bent2008dynamic,
zhang2022skeena, microsoft-fabric} also aim to distribute query workloads across
heterogeneous engines.
Unlike \thesystem{}, these systems focus on
\begin{enumerate*}[label=(\roman*)]
  \item optimizing queries within a given set of engines and hardware configuration,
    and
  \item bridging different data models~\cite{duggan15bigdawg}.
\end{enumerate*}
In contrast, \thesystem{} tackles the problem of selecting the best set of
engines to include in the underlying infrastructure for the user's workload
(among Aurora, Redshift, and Athena), while also jointly optimizing the
workload assignment, engine provisioning, and data placement.

\section{Conclusion}
This paper presents \emph{blueprints}, \emph{blueprint planning}, and
\thesystem{}: a system that virtualizes a cloud data infrastructure and
leverages blueprint planning to automatically manage its physical realization.
The key takeaway is to cast infrastructure design as a \emph{cost-based
optimization problem}, which we refer to as \emph{blueprint planning}.
This approach allows us to systematically search for an optimized design for a
given workload by leveraging learned models to predict the utility of candidate
blueprints.
We show that \thesystem{} automatically achieves performance targets while
saving \BRADCostSavingsRange{} in cost compared to existing serverless
autoscaling systems.

\begin{acks}
This research was supported by Amazon, Google, and Intel as part of the MIT Data
Systems and AI Lab (DSAIL) at MIT and NSF IIS 1900933. Geoffrey X. Yu was
partially supported by an NSERC PGS D. This research was also sponsored by the
United States Air Force Research Laboratory and the Department of the Air Force
Artificial Intelligence Accelerator and was accomplished under Cooperative
Agreement Number FA8750-19-2-1000. The views and conclusions contained in this
document are those of the authors and should not be interpreted as representing
the official policies, either expressed or implied, of the Department of the Air
Force or the U.S. Government. The U.S. Government is authorized to reproduce and
distribute reprints for Government purposes notwithstanding any copyright
notation herein.
\end{acks}

\bibliographystyle{ACM-Reference-Format}
\bibliography{brad}

\end{document}